\documentstyle[preprint,aps]{revtex}
\newcommand\lsim{\, 
  \raisebox{-.5ex}{$\stackrel{<}{\scriptstyle\sim}$} \, }
\newcommand\gsim{\, 
   \raisebox{-.5ex}{$\stackrel{>}{\scriptstyle\sim}$}\, }
\begin{document}
\draft
\preprint{HU--TFT--96--08; JYFL 96-9;
SUNY-NTG-96-25; DOE/ER/40561--247--INT96-00-116}
\title{ Hydrodynamical Description of $200\ A$ GeV/$c$ S+Au
Collisions:\\
 Hadron and Electromagnetic Spectra}
\author {Josef Sollfrank}
\address{Research Institute for Theoretical Physics, 
 University of Helsinki, Finland \\ }
\author { Pasi Huovinen, Markku Kataja, P.V.~Ruuskanen}
\address{Department of Physics,  
  University of Jyv\"askyl\"a, Finland \\ }
\author {Madappa Prakash}
\address{Physics Department, SUNY at Stony Brook,
Stony Brook, USA \\}
\author{Raju Venugopalan}
\address{National Institute for Nuclear Theory, 
 University of Washington, Seattle, USA \\ }
\date{\today}
\maketitle
\begin{abstract}
We study relativistic S+Au collisions at $200 A$ GeV/$c$ using a
hydrodynamical approach.  We test various equations of state
(EOSs), which are used to describe the strongly interacting matter
at densities attainable in the CERN-SPS heavy ion experiments.
For each EOS, suitable initial conditions can be determined to
reproduce the experimental hadron spectra; this emphasizes the
ambiguity between the initial conditions and the EOS in such an
approach.  Simultaneously, we calculate the resulting thermal photon
and dielectron spectra, and compare with experiments. If one allows
the excitation of resonance states with increasing temperature, the
electro-magnetic signals from scenarios with and without phase
transition are very similar and are not resolvable within the current
experimental resolution. Only EOSs with a few degrees of freedom up to
very high temperatures can be ruled out presently. We deduce an upper
bound of about 250 MeV for the initial temperature from the single
photon spectra of WA80.  With regard to the CERES dilepton data, none
of the EOSs considered, in conjunction with the standard leading order
dilepton rates, succeed in reproducing the observed excess of
dileptons below the $\rho$ peak.  Our work, however, suggests that an
improved measurement of the photon and dilepton spectra has the
potential to strongly constrain the EOS.
\end{abstract}
\vspace*{0.2in} 

%%%%%%%%%%%%%%%%%%%%%%%%%%%%%%%%%%%%%%%%%%%%%%%%%%%%%%%%%%%%%%%%%
\section{Introduction}
%%%%%%%%%%%%%%%%%%%%%%%%%%%%%%%%%%%%%%%%%%%%%%%%%%%%%%%%%%%%%%%%%

The use of hydrodynamics for simulations of nuclear collisions has a
long tradition \cite{Landau53,Stoecker86,Clare86}.  If applicable,
hydrodynamics has some advantages over the more fundamental kinetic
calculations, which are usually performed as Monte-Carlo simulations.
Besides its relative simplicity, the use of familiar concepts such as
temperature, flow velocity, energy density, pressure, etc.~leads to an
intuitively transparent picture of the space-time evolution of the
system.  Another great advantage is the direct use of the equation of
state (EOS) of strongly interacting matter.  Testing different EOSs by
comparing with experiments should give us more insight into the
behavior of nuclear matter under extreme conditions of temperature
and/or density.  This is important, since one of the main reasons to
study high energy heavy ion collisions is to confirm the possible
phase transition from hadronic matter to the quark-gluon plasma (QGP)
\cite{Hwa91}.  The necessary energy densities are estimated to be
around 1--2 GeV/fm$^3$, presently experimentally available at the
Brookhaven AGS and the CERN-SPS.  Of these two facilities, conditions
to observe signals of the phase transition are more favorable at the
CERN-SPS, because of the higher incident momenta of the nuclei.

\vspace*{0.2in}

In this work, we analyze the data from the heavy ion experiments at
the CERN-SPS with $200 A$ GeV/$c$ incident momentum
\cite{Alber94}--\cite{Agakichiev95}, concentrating on the S + Au
system. For this system, yields of both the hadronic and
electromagnetic (dileptons and photons) probes are now available. A
hydrodynamical treatment of the nuclear collisions in this energy
range is not new \cite{Venugopalan94}--\cite{Katscher93}, but the
large set of new and updated experimental data allows us to achieve a
deeper insight into the space-time evolution of these reactions.

\vspace*{0.2in}

Our emphasis here is on the simultaneous description of the hadron and
electromagnetic data.  Since hadrons interact throughout the dense
phase of the collision, the hadronic spectra are chiefly determined
from the conditions at the freeze-out of matter.  However, different
initial conditions combined with different EOSs can lead to the same
final particle distributions.  In contrast, photon and lepton pairs
are emitted throughout the dense stage and escape without
interactions.  The measured spectra thus probe the different
temperature and flow conditions during the evolution.  These should
have a different dependence on the initial conditions and on the EOS,
in comparison to the hadron spectra.  This may help to further reduce
the ambiguity in the initial conditions and uncertainties in the EOS.

\vspace*{0.2in}

Our hydrodynamical calculations assume azimuthal symmetry around the
beam axis.  Thus, we simulate only central collisions with impact
parameters close to zero.  We do not describe the production of matter
within hydrodynamics, but start the calculation at an initial time
when the system is likely to have reached thermal and chemical
equilibrium.  We explore different initial conditions, taking guidance
from the experimental data (where available) on particle production in
proton-proton $(p+p)$ collisions in choosing what could be considered
as a realistic initial state.  However, one must be aware that nuclear
effects are likely already presentat the production stage, and the
initial conditions cannot be uniquely determined from $p+p$ processes
alone.

\vspace*{0.2in}

A major concern is the uncertainty in the nuclear EOS.  It is
experimentally and theoretically well known only around the ground
state of nuclear matter.  At higher temperatures and densities, model
predictions are widely different.  The main goal in this study is to
derive constraints on the EOS by comparing the hydrodynamical
calculations with hadronic and electromagnetic data simultaneously
from S + Au collisions at $200 A$ GeV/$c$.

\vspace*{0.2in}

The paper is organized as follows: In Sec II,
%\ref{hydroequation}
we briefly introduce the hydrodynamical equations and address some
problems which arise when seeking numerical solutions of these
equations. In Sec. III,
%\ref{seceos},
the different EOSs explored are described.  The initial conditions are
discussed in Sec. IV.
%\ref{initialcondition}.
In Secs. V, VI and VII, we describe the calculation of the final
hadron, photon, and dilepton spectra, respectively. Results are
discussed in Sec. VIII,
%\ref{results}
and our conclusions are given in Sec. IX. In the appendix, we briefly
explain how we incorporated the CERES kinematic cuts and the detector
resolution.

%%%%%%%%%%%%%%%%%%%%%%%%%%%%%%%%%%%%%%%%%%%%%%%%%%%%%%%%%%%%%%%%%
\section{The Hydrodynamical Equations}
\label{hydroequation}
%%%%%%%%%%%%%%%%%%%%%%%%%%%%%%%%%%%%%%%%%%%%%%%%%%%%%%%%%%%%%%%%%

The basic equations of hydrodynamics are the local conservation of
energy and momentum, which in Lorentz-covariant form are written as
\begin{equation}\label{relhydro}
\partial_\mu T^{\mu\nu} (x) = 0 \: .
\end{equation}
We use the ideal fluid ansatz for the energy-momentum tensor
\begin{equation}\label{deft}
T^{\mu\nu}(x) = [\varepsilon(x) + p(x)]u^{\mu}(x)u^{\nu}(x) -
p(x)g^{\mu\nu} \,,
\end{equation}
where $\varepsilon(x)$ is the local energy density, $p(x)$ is the
local pressure, and $u^{\mu}(x)$ is the local four-velocity,
normalized to $u^{\mu}u_{\mu} = 1$.  In principle, viscous effects may
be included, but would lead to a major increase in the numerical
effort \cite{Clare86,Gersdorff86}. However, the calculation generates
some numerical viscosity, as explained below.

\vspace*{0.2in}

We include finite baryon density, $\rho_{\rm B}$, in our system and
express baryon number conservation locally in the form
\begin{equation}\label{relb}
\partial_\mu j^{\mu}_{\rm B} = 0 \: ,
\end{equation}
in terms of the baryon current $j^{\mu}_{\rm B}= \rho_{\rm B}u^{\mu}$.

Using the definition Eq. (\ref{deft}), the continuity
Eqs.~(\ref{relhydro}) and (\ref{relb}) can be written more explicitly
as
\begin{eqnarray}
\partial_t T^{00} + \vec{\nabla}\cdot(T^{00}\vec{v}) & = & -
\vec{\nabla}\cdot(p\vec{v}) \, , \nonumber \\ \partial_t T^{0i} +
\vec{\nabla}\cdot(T^{0i}\vec{v}) & = & - \partial_i p \, , \nonumber
\\ \partial_t j_{\rm B}^{0} + \vec{\nabla}\cdot(j_{\rm B}^{0}\vec{v})
& = & 0 \,.
\label{partdif}
\end{eqnarray}
In order to solve Eq.~(\ref{partdif}), one additional input is needed,
namely, the equation of state (EOS).  The EOS relates the pressure
$p=p(\varepsilon,\rho_{\rm B})$ to the energy and baryon
densities. The different choices of the EOS are described in the next
section.

\vspace*{0.2in}

The partial (hyperbolic) differential equations, Eq.~(\ref{partdif}),
are solved numerically on a computer using a finite difference method.
We use the SHASTA algorithm \cite{Boris73} in two spatial dimensions,
radial and longitudinal.  Some details of implementing this algorithm
are explained in the appendix of Ref.  \cite{Gersdorff86}.  However,
there are two essential modifications.  Firstly, since we now have a
problem with two spatial dimensions \cite{Kataja88}, the flux
correction must be modified accordingly, as described in
Ref. \cite{Zalesak79}.  Secondly, the flux correction involves a
parameter $\eta$ \cite{Gersdorff86,Rischke95a}, the anti-diffusive
constant, which regulates the residual numerical anti--diffusive flux
in the algorithm.  Theoretically, for $\eta = 1/8$, diffusion vanishes
almost completely, leading, however, to the appearance of ripples in
the presence of steep gradients.

\vspace*{0.2in}

On the other hand, a residual diffusive flux mimics viscosity in the
calculation, since it generates entropy in the system. In
Ref. \cite{Gersdorff86}, a large residual diffusive flux was allowed
by choosing a small value for the anti-diffusive constant,
$\eta=0.08$.  This value was determined by trying to create as much
entropy as is maximally allowed in the rarefaction wave, typically
$\lsim 10\%$.  However, it turns out that the flux across the
freeze-out surface (Eq.~(\ref{cf}) below) now has two contributions;
the normal convective part and the diffusive part. Because we cannot
assign a velocity to the numerical diffusive flux, it cannot be
converted into particle spectra at the freeze-out surface. Therefore,
we diminish the diffusive flux by setting $\eta = 0.11$, as suggested
also in Ref. \cite{Rischke95a}, and neglect the small residual
diffusive flux across the freeze-out hypersurface. This leads to a
small, of order $\lsim 5$\%, loss in the baryon number and total
energy across the freeze-out hypersurface.  (The stability of the
numerical calculation increases as $\eta$ is diminished from the
limiting value $1/8=0.125$. Our choice, $\eta=0.11$, is a compromise
between numerical stability and the diffusive loss across the
freeze-out surface.)  The value of $\eta$ has a small influence on the
space-time evolution, a larger diffusive flux leading to a more rapid
cooling of the system.  This has only a small effect, especially on
the electromagnetic spectra.

\vspace*{0.2in}

At the moment, our numerical code assumes a mirror symmetry in the
longitudinal direction with respect to the center of mass ($f(z) =
f(-z)$ for scalar quantities). This is a very good approximation for
symmetric colliding systems.  However, we apply it here to the
asymmetric S + Au collisions.  Thus, we are presently unable to
reproduce the asymmetry in the rapidity distributions of hadrons,
which is observed in experiment. However, in the central rapidity
region, where most of the experimental data are measured, we do not
expect this to affect significantly the transverse spectra or the
electromagnetic yields.  We will consider the longitudinal asymmetry
in a later work.

%%%%%%%%%%%%%%%%%%%%%%%%%%%%%%%%%%%%%%%%%%%%%%%%%%%%%%%%%%%%%%%%%%
\section{The Equation of State}
%\label{seceos}
%%%%%%%%%%%%%%%%%%%%%%%%%%%%%%%%%%%%%%%%%%%%%%%%%%%%%%%%%%%%%%%%%%

The equation of state (EOS) in the energy density domain of
$\varepsilon \gsim 1$ GeV/fm$^3$ or baryon density $\rho_{\rm B} \gsim
2\rho_0$, where $\rho_0$ is the density of ground state nuclear
matter, is theoretically quite uncertain.  At the CERN-SPS energies,
the relevant energy densities are not high enough that the well
established techniques of finite temperature {\it perturbative} QCD
will apply. At the same time, the energy densities are typically too
high for low energy approaches, such as chiral perturbation theory, to
be applicable.  Hadronic and quark--gluon matter at these temperatures
and densities, and especially the chiral and/or de-confinement
transitions between the two phases at some critical temperature and
density, are non--perturbative phenomena which are currently
investigated through lattice simulations of QCD.

\vspace*{0.2in}

Simulations of pure SU(3) gauge theory indicate the occurence of a
first-order deconfinement transition around $T_c \simeq 260$ MeV
through studies of hysterisis, co--existing states, and abrupt
quantitative changes in the various thermodynamic functions. Recent
simulations on larger lattices (16$^3$ $\times$ 32) suggest that
finite size effects are reasonably well understood for pure glue ---
an estimate for the critical energy density for pure SU(3) is $\sim 1$
GeV/fm$^3$\cite{karschQM95}.  For full lattice QCD with dynamical
quarks, the situation is less clear, particularly in the critical
region.  However, a recent calculation~\cite{Blum95} shows that the
energy density displays a rapid change in a narrow region around
$T_c\simeq 150$ MeV, while the pressure changes more smoothly.

\vspace*{0.2in}

Though the present uncertainities preclude a quantitative description
of the thermodynamic functions around the critical region, one may
observe i) a smooth crossover between hadronic matter and the quark
gluon phase and ii) a sizable and rapid change in the entropy
density. These features may be roughly reproduced in a two phase
description of the transition. Indeed, even if the transition were a
{\em sharp} first order transition, it is unlikely that the
hydrodynamic flow simulations would be extremely sensitive to the
width (in temperature) of the critical
region~\cite{BlaiOll86}. Therefore, for the purpose of our
hydrodynamic simulations, we will construct EOSs for both the hadron
and quark gluon state variables and match them at the critical
boundary in temperature and chemical potential by a Maxwell
construction. In practice, when implemented in simulations, this
boundary is smoothed over using a hyperbolic tangent profile of width
$\Delta$. It has been checked that the results of our simulations are
insensitive to $\Delta$, as long as $\Delta/T_c << 1$
~\cite{RajThesis}. In addition to studying EOSs with a sharp cross
over (but differing critical temperatures and latent heat), we will
also consider an EOS with purely hadronic degrees of freedom.

\vspace*{0.2in}

Let us first discuss the EOS we use for the hadronic phase. For a
dilute hadronic gas at temperatures well below the pion mass, state
variables can be computed reliably using a virial expansion with input
from empirical scattering cross sections~\cite{WVP,Venugopalan92}. It
was observed that the state variables for an interacting hadron gas
are well approximated by those for a Boltzmann gas of free hadrons and
resonances. At temperatures comparable to the pion mass or higher,
third and higher virial coefficients are important. Presently, there
is no reliable way to compute these. For want of a systematic
prescription, we shall stretch our conclusions from the virial
expansion approach and assume that a dense hadron gas is roughly
approximated by a gas of free hadrons and resonances.

\vspace*{0.2in}

We restrict our studies here to a hadron gas, which contains the
following hadrons with their corresponding anti-particles:
\begin{equation}\label{hadroncomp}
\pi, K, \eta, \rho, \omega, K^*, p, n, \eta^{\prime}, \phi, \Lambda,
\Sigma, \Delta, a_1, \Xi, \Sigma(1385).
\end{equation}
Beside stable particles (on strong interaction time scales), the
hadron gas also contains resonance states.  This should mimic the
attractive interaction among the hadrons in the spirit of the
bootstrap model of Hagedorn \cite{Hagedorn94}, although we keep the
number of resonance states finite.  If we assume a phase transition to
the QGP at temperatures around $T_c \approx 150$ MeV, then the limited
number of hadronic states is justified, because of the suppression of
higher mass states through the Boltzmann factor.

\vspace*{0.2in}

As mentioned above, we will also consider a pure hadronic EOS, which
does not exhibit a phase transition.  In this case, somewhat higher
temperatures are encountered in our calculations.  Therefore, a
sensitivity on the upper mass cut in the hadronic mass spectrum could
be expected. We expect this to show up more in the values of the
temperature rather than in the evolution of the flow. While hadronic
spectra are not very sensitive to this cut, electromagnetic rates,
however, could decrease. On the other hand, for consistency, we should
include the increase in electromagnetic emission from processes
involving the heavy resonances. To some extent, these effects tend to
cancel each other.  Mainly due to the numerical limitations of our
hydrodynamical simulations, we have included only those states listed
above, even at high temperatures.

\vspace*{0.2in}

In order to derive the EOS, we begin with the grand canonical
partition function for a non-interacting resonance gas. It is given by
\begin{equation}
 \label{zhad} {\cal Z}^{\rm H} (T, V, \mu_{\rm B}, \mu_{\rm S}) =
   \prod\limits_h \exp\left[ Z_h(T,V,\mu_h) \right]\: ,
\end{equation}
where $V$ is the volume, and the product is over the different hadron
species $h$. The chemical potential $\mu_h$ of the hadron $h$ is given
by its baryon number $B_h$ and its strangeness $S_h$ through
\begin{equation}\label{muh}
\mu_h = B_h \mu_{\rm B} + S_h \mu_{\rm S}
\end{equation}
in full chemical equilibrium. We will use this assumption throughout
the expansion, although there are indications of deviations,
especially for the strange particles \cite{Letessier95}.  The
partition function Eq.~(\ref{zhad}) is, in general, a function of four
unknown variables. For calculations of intensive variables, the volume
cancels and the value of the strangeness chemical potential $\mu_{\rm
S}$ can be related to $T$ and $\mu_{\rm B}$ by the requirement of
local strangeness neutrality.

The partition function $Z_h(T,V,\mu_h)$ for hadron species $h$ is
\begin{equation}
 \label{zone} \ln Z_h(T,V,\mu_h) = \beta Vp_h = \frac{g_h \beta
   V}{6\pi^2} \int\limits_{m_h}^{\infty} {\rm d} E \: \frac{(E^2 -
   m_h^2)^{3/2}}{ e^{\beta(E-\mu_h)} \pm 1} \: ,
\end{equation}
where $g_h$ is the degeneracy factor, $m_h$ is the mass, and $\mu_h$
is the chemical potential of hadron $h$. The $\pm$ sign corresponds to
fermions and bosons, respectively. From the partition function
Eq.~(\ref{zhad}), we can calculate all thermodynamical quantities.
Specifically, we have
\begin{eqnarray}
\label{bdensdef}
   \rho_{\rm B}(T, \mu_{\rm B}) & = & \frac{T}{V} \: \frac{\partial
   \ln{\cal Z}^{\rm H}}{\partial \mu_B}\\
\label{edensdef}
   \varepsilon(T, \mu_{\rm B}) & = & \frac{1}{V} \: \frac{\partial
   \ln{\cal Z}^{\rm H}}{\partial \beta} \\
\label{thermqant}
p (T, \mu_{\rm B}) & = & T \: \frac{\partial \ln{\cal Z}^{\rm
        H}}{\partial V} = \frac{T}{V} \: \ln{\cal Z}^{\rm H}\,.
\end{eqnarray}
From these equations, the pressure can be solved as a function of the
energy density $\varepsilon$ and the baryon density $\rho_{\rm
B}$. This form of the EOS, $p=p(\varepsilon,\rho_{\rm B})$, is needed
in solving the hydrodynamic equations and in practice is obtained
numerically from Eqs.~(\ref{edensdef}) and (\ref{thermqant}).

\vspace*{0.2in}

In the limit of high baryon densities, the repulsive interactions
between particles have to be taken into account. Otherwise, the
hadronic phase is preferred over the quark-gluon phase.  At the
temperatures of interest, repulsive interactions reduce the
contributions from the high mass part of the spectrum
\cite{Kapusta89}, justifying our use of a finite number of resonance
states in the EOS.  It has been shown \cite{Venugopalan92} that within
the different ways to include repulsion, a mean field approach, such
as in Ref. \cite{Kapusta83a}, gives the most realistic results.  Thus,
we will introduce a repulsive mean field potential ${\cal V}$, which
we couple only to the net baryon density.  This is similar in spirit
to the Walecka model \cite{Walecka74}. By including many of the
resonance states, the main attractive interactions (akin to the scalar
interactions of the Walecka model) are already taken into account.

\vspace*{0.2in}

We assume that the repulsive potential energy density ${\cal V}$ is a
function of $\rho_{\rm B}$ of the form
\begin{equation}\label{reppot}
{\cal V}(\rho_{\rm B}) =\frac{1}{2}\, K\rho_{\rm B}^2\, ,
\end{equation}
where $K$ is the mean field repulsion parameter.  The partition
function ${\cal Z}^{\rm MF}$ with a mean field interaction
Eq. (\ref{reppot}) is given by \cite{Greiner}
\begin{equation}
 \label{zmf} {\cal Z}^{\rm MF} = \exp\left\{-\beta V[{\cal
   V}(\rho_{\rm B}) - \rho_{\rm B}{\cal V}^{\prime}(\rho_{\rm B})]
   \right\} \prod\limits_h \exp\left[ Z_h(T,V,\mu_{\rm B}^{\rm
   eff},\mu_{\rm S}) \right]\: ,
\end{equation}
where the effective baryon chemical potential
\begin{equation}
\mu_{\rm B}^{\rm eff} = \mu_{\rm B} - {\cal V}^{\prime}(\rho_{\rm B})
= \mu_{\rm B} - K\rho_{\rm B}\:
\end{equation}
describes the shift in the particle energy by $K\rho_{\rm B}$ due to
the repulsive interaction. Using Eq. (\ref{bdensdef}) with
Eq. (\ref{zmf}) leads to a self-consistency equation for the baryon
density
\begin{equation}
\rho_{\rm B} = \sum\limits_{b} \int d^3 p \: \frac{1}{\exp{[(E -
               \mu_{\rm B}B_b + K\rho_{\rm B} - \mu_{\rm S}S_b)/T]} +
               1} \: .
\end{equation}
Once $\rho_{\rm B}$ is solved, applying Eqs.~(\ref{edensdef}) and
(\ref{thermqant}) to the mean field partition function,
Eq. (\ref{zmf}), gives the repulsion corrected energy density and
pressure. The resulting EOS is labeled as EOS~H and has one free
parameter $K$. The ideal gas equation of state $p(\varepsilon,
\rho_{\rm B}) = \varepsilon/3$, valid for massless non-interacting
particles, is referred to as EOS~I.  We use this with three massless
pions, to illustrate the effect of the number of hadronic degrees of
freedom.

\vspace*{0.2in}

So far, we have considered the equation of state using only the
hadronic degrees of freedom. We will now construct an EOS having a
phase transition to the QGP. As discussed earlier, the QGP in the
critical region is highly nonperturbative and is best understood from
QCD lattice simulations. For simplicity, the results of these
simulations can be parametrized in terms of an ideal massless parton
gas. In terms of temperature $T$ and chemical potential $\mu_B$, the
thermodynamic densities are given by
\begin{eqnarray}
p & = & \frac{(32 + 21 N_f)\pi^2}{180} \: T^4 + \frac{N_f}{2} \:
       \mu^2_{\rm q} \: T^2 + \frac{N_f}{4\pi^2} \: \mu^4_{\rm q} - B
       \\ \varepsilon & = & 3p+4B \\ \rho_{\rm B} & = & N_f \:
       \mu^2_{\rm q} \: T + \frac{N_f}{\pi^2} \: \mu^3_{\rm q} \: ,
\end{eqnarray}
where $\mu_{\rm q} = \mu_{\rm B}/3$.  $B$ is the bag constant, and we
use the number of flavors $N_f = 2.5$ in order to simulate effects of
the finite strange quark mass.

\vspace*{0.2in}

The phase boundary is determined by the pressure balance, $p_{\rm HG}
= p_{\rm QGP}$, between the two phases at equilibrium.  In the mixed
phase, $\varepsilon$ and $\rho_{\rm B}$ are calculated using the
Maxwell construction. We define
\begin{equation}\label{defw}
w(\varepsilon,\rho_{\rm B}) = \frac{V^{\rm QGP}} {V^{\rm HG}+V^{\rm
QGP}}\;\; ; \;\; 0 \le w \le 1
\end{equation}
as the volume fraction of the QGP in the mixed phase.

\vspace*{0.2in}

The resulting EOS depends on two parameters, the mean field repulsion
parameter $K$ and the bag constant $B$. In this work, we give results
for only two different choices of $K$ and $B$, and call these
parametrizations EOS~A and EOS~B. The two parameter sets are chosen to
resemble the known features of the phase transition from lattice
calculations \cite{Blum95}, the difference encompassing the
uncertainty of lattice results.  These indicate a value $T_c \approx
140 - 160$ MeV for the transition temperature and $\Delta \varepsilon
\approx 1 - 1.5$ GeV/fm$^3$ for the latent heat.  Our EOS~A, given by
$K = 450$ MeV fm$^3$ and $B^{1/4} = 235$ MeV, is similar to the one
used in Ref. \cite{Rischke95b}.  The resulting $T_c$ is 165 MeV, and
$\Delta \varepsilon = 1.4$ GeV/fm$^3$.  For EOS~B, the parameters are
$K = 660$ MeV fm$^3$ and $B^{1/4} = 200$ MeV, resulting in $T_c = 140$
MeV and $\Delta \varepsilon = 0.8$ GeV/fm$^3$.  This parametrization
should represent reasonable lower bounds for $T_c$, $\mu_c$, and the
latent heat. The parameters for the EOSs we use are summarized in
Table \ref{tab1}.

\vspace*{0.2in}

The phase diagrams of equations of state A and B are shown in
Fig.~\ref{phaseeos}. The difference in the latent heat can be seen if
one plots the phase boundaries in the $\rho_{\rm B}$--$\varepsilon$
plane. This can be read off from Fig.~\ref{eosfigure}, where the
pressure is shown as a function of $\rho_{\rm B}$ and $\varepsilon$.
For the EOS~A, there is a large increase in the pressure, even in the
mixed phase, with increasing baryon density, while for the EOS~B the
increase is much smaller.  It would be interesting to obtain from
hydrodynamical calculations constraints on the EOS at finite baryon
density, since nothing is known so far from lattice calculations in
this region. Presently, however, the large uncertainties in the
initial conditions preclude definitive conclusions to be drawn.

%%%%%%%%%%%%%%%%%%%%%%%%%%%%%%%%%%%%%%%%%%%%%%%%%%%%%%%%%%%%%%
\section{The Initial Conditions}
\label{initialcondition}
%%%%%%%%%%%%%%%%%%%%%%%%%%%%%%%%%%%%%%%%%%%%%%%%%%%%%%%%%%%%%%

Hydrodynamical simulations at energies near or below 10 $A$ GeV
(BEVALAC and AGS) usually start with the colliding nuclei before the
impact and include the initial compression and particle production.
In the one-fluid hydrodynamics, the nuclei fuse to a single fluid,
implying, at zero impact parameter, a complete stopping of equal size
nuclei.  At higher energies, as at the CERN-SPS, RHIC and LHC, this is
not justified, and one must be able to incorporate nuclear
transparency in the description.  Instead of trying to describe the
production and equilibration within hydrodynamics, we start the
calculation from initial conditions which specify the hydrodynamic
state of the system at time $t_0$.  Initial conditions parametrize the
production and equilibration dynamics.  Note that $t_0$ only plays the
role of a bookkeeping device in the numerical calculations.
Thermalization time does not enter explicitly in the parametrization,
but, physically, it may be related to the initial longitudinal size,
$z_0$, at thermalization. In some cases, we find it useful to relate
$z_0$ with $t_0$ when defining the initial longitudinal velocity
profile.

\vspace*{0.2in}

Physically, the uncertainty in the initial conditions arises mainly
from the lack of definitive knowledge about the nuclear stopping power
and the time scale for equilibration.  Two extreme scenarios are the
full stopping model of Landau \cite{Landau53} and the full scaling
expansion model espoused by Bjorken \cite{Bjorken83}, in which
$v_z=z/t$, beginning with equilibration.  Although the precise
energies at which these extreme cases are practically useful is
unknown, the first case is expected to apply at moderately high
energies and the second at ultrarelativistic energies.  Since the
present experiments fall between these limits, we have developed
parameterizations which span the range between these two extremes. We
will try to incorporate some known features of hadron-hadron
collisions in the parametrization, which are also constrained to
satisfy the conservation of energy and baryon number.

\vspace*{0.2in}

We first consider the initial velocities.  The collective four
velocity is denoted by $u^\mu = \gamma(1,\vec{v})$, where $\vec{v}$ is
the flow velocity vector.  Since we consider zero impact parameter
cylindrically symmetric collisions only, we do not expect significant
collective motion initially, and take the velocity in the radial
direction at $t_0$ to vanish, i.e., $\vec v_{\rm r}(t_0,r)=0$.  Note
that for the Landau initial conditions, the initial longitudinal
velocity also vanishes, i.e., $v_z(t_0,r,z)=0$.  Strictly speaking,
the Bjorken model applies only in the infinite energy limit.  In this
case, the scaling ansatz for the four velocity is
\begin{equation}\label{vbj}
u^{\mu}(z,t) = \frac{1}{\tau}(t,0,0,z) \,,
\end{equation}
implying $v_z = z/t$.  Invariance under longitudinal Lorentz boosts
means that the thermodynamic quantities depend only on the
longitudinal proper time $\tau = \sqrt{t^2 - z^2}$, which equals the
local time in the rest frame of any fluid element.

\vspace*{0.2in}

At finite, albeit high collision energies, the longitudinal extent of
the system is finite and the Bjorken scenario does not work properly
in the fragmentation regions.  Since we perform the numerical
calculation using variables in the center-of-mass frame of the
participating nucleons, we specify the initial condition at a fixed
time $t_0$ in this frame. We choose an initial velocity profile of the
following form:
\begin{equation}\label{vap}
v_z(z) = \tanh(z/t_0)\,,\qquad y(z) = z/t_0 \:.
\end{equation}
In this form, $t_0$ should be regarded as a constant, which fixes the
rapidity of produced matter at the edge $z_0$, rather than the
equilibration time.

\vspace*{0.2in}

The reason for taking the rapidity $y$, instead of velocity $v_z$, to
be proportional to $z$ is purely practical. For numerical
calculations, initial conditions must be smoothed and extended over
the edge of the produced matter, initially at $z_0$. The above
parametrization leads to the Bjorken limit, Eq. (\ref{vbj}), in the
inner part ($z/t_0 \ll 1$), and extrapolates the velocity smoothly to
unity in the outer parts, where the thermodynamic densities approach
zero.

\vspace*{0.2in}

The energy density distribution in the Bjorken model is constant along
contours of equal proper time $\tau_0(t,z) = \sqrt{t^2-z^2}$.
However, in the global frame, at fixed time $t=t_0$, it scales with
$z$ as ($\varepsilon_0 := \varepsilon(t_0,0)$) \cite{Bjorken83}
\begin{equation}\label{sbj}
\frac{\varepsilon(t_0,z)}{\varepsilon_0} =
\left(\frac{t_0}{\sqrt{t_0^2-z^2}}\right)^{4/3} = \gamma(z)^{4/3}\,,
\end{equation}
if an ideal gas EOS, $\varepsilon = 3p$, is assumed. In the last term,
$\gamma(z) = (1-(z\slash t_0)^2)^{-1\slash 2}$ is the relativistic
$\gamma$--factor for the Bjorken expansion velocity, Eq.~(\ref{vbj}).

\vspace*{0.2in}

We use Eq.~(\ref{sbj}) for our initial parametrization, despite the
fact that it is strictly valid only for the velocity field
Eq. (\ref{vbj}), i.e., the Bjorken picture.  One factor of $\gamma$ in
this equation expresses the time dilatation effect for the moving
cells and the factor $\gamma^{1/3}$ the energy loss due to the work
done by pressure in the expansion. Retaining this factor whenever the
initial longitudinal velocity is non-vanishing is therefore physically
well motivated.  Then the remaining factor in $\varepsilon(t_0,z)$
represents the energy density in the local rest frame, and values at
different $z$ have the same physical interpretation.  In the Bjorken
picture, a rapidity plateau indicates constant local energy
density. At fixed global initial time, this leads to increasing
$\varepsilon(t_0,z)$ with increasing $z$. Since, at finite collision
energy, the thermodynamic densities vanish when $z > z_0$, this
constraint will ensure that the energy density will have a maximum at
finite $z$ .  This is in contrast to Ref.~\cite{Bolz92}, in which the
authors employed a a constant energy density along $z$ until the edges
are reached.

\vspace*{0.2in}

In production processes, distributions are always cut off smoothly
when the phase space boundary is approached. Smoothing also helps to
avoid oscillations in the numerical calculations. We implement
smoothing by multiplying the distributions with a Fermi function
\begin{equation}
f(x,x_0,a_x) = \frac{1}{\exp[(|x| - x_0)/a_x] + 1} \: ,
\end{equation}
where $a_{x},\ x=r,z$ is the diffuseness parameter in the radial and
the longitudinal directions, respectively. The initial energy
distribution is then given by
\begin{equation}\label{energydist}
\varepsilon(z,r) = \varepsilon_0\:\gamma(z)^{4/3} \: f(z,z_0,a_z) \:
f(r,r_0,a_r) \: .
\end{equation}
Since our code is symmetric in $z$ in the center-of-mass frame of the
participant nucleons, or equivalently, of the produced fireball, we
cannot reproduce the asymmetric shape of rapidity distributions in
S+Au collisions.  In the central region, $y=0$, we expect effects from
the asymmetry on the expansion dynamics to be less important than
those from the average longitudinal gradients, which are properly
accounted for in the code.

\vspace*{0.2in}

In the case of the longitudinal scaling expansion, the rapidity
density of baryon number, $dN_{\rm B}\slash dy$ does not change during
the expansion.  This allows us to relate the rapidity density to the
initial baryon density through the equation
\begin{equation}\label{b1}
\rho_{\rm B}(z) = \frac{1}{\pi R_{\rm proj}^2 dz \gamma(z)}\:
                   \frac{dN_{\rm B}}{dy} \: \frac{dy}{dz} \: dz=
                   \frac{1}{\pi R_{\rm proj}^2 \tau}\: \frac{dN_{\rm
                   B}}{dy} \: \left[\frac{\tau}{\gamma(z)}
                   \frac{dy}{dz}\right]\,,
\end{equation}
where, for the scaling expansion, the expression in square brackets
equals unity.  In our case, expansion modifies the rapidity
distributions and the problem is to find a reasonable choice for the
initial $dN_{\rm B}/dy$.

\vspace*{0.2in}

In nucleon-nucleon collisions, the leading particle effect in the
nucleon distributions causes more nucleons to be present in the
fragmentation regions than at central rapidity
\cite{Blobel74,Otterlund83}.  In nuclear collisions, however, more
stopping is present and it depends on the mass numbers of the
colliding nuclei.  Here, we assume that the initial rapidity density
of baryons is flat, implying constant density at constant proper time
in the Bjorken parametrization.  At fixed global time, the baryon
distribution as a function of $z$ is then
\begin{equation}\label{b2}
\rho_{\rm B}(z) = b_0 \: \gamma(z) \:f(z,z_0,a_z) \: f(r,r_0,a_r) \: ,
\end{equation}
where $b_0$ is the central baryon density.

\vspace*{0.2in}

Some of the initial parameters can be fixed by geometry. A natural
choice for $r_0$ is the sulphur radius, 3.65 fm. For the diffuseness
parameter in a Woods-Saxon \cite{Woods54} parametrization of the
nuclear surface, we take the value $a_r = 0.5$ fm. We set $a_z = 0.13$
fm in order to have $a_r/r_0 \approx a_z/z_0$. The number of
participating nucleons is calculated from geometry, by assuming that
at impact parameter zero the sulphur nucleus coalesces with a central
tube of the gold nucleus having the diameter of sulphur. From the
calculated value $B_{\rm tot}= 104$, we obtain the rapidity of the cm
frame of the participating nucleons, $y_{\rm cm} = 2.6$, and the total
cm energy, $E_{\rm tot}=970$ GeV. This total energy should be
considered to give the overall scale of the energy in the fireball,
since small changes in the rapidity distributions of final particles
in the fragmentation regions easily cause 10-20\% changes in the total
energy. We do not consider such deviations to be significant when
fitting the pion spectra. Given $E_{\rm tot}$ and $B_{\rm tot}$, we
can fix two of the remaining five parameters in the initial
conditions. These are the central initial energy density
$\varepsilon_0$ and baryon density $b_0$.

\vspace*{0.2in}

The real ambiguity in the initial condition lies in the parameters
$z_0$ and $t_0$.  While in the Bjorken scenario, they are related
through $t_0 = z_0/c$, i.e., the edge of matter is by definition on
the light cone; this does not hold in a finite energy scenario, like
the one introduced here.  Now, the initial longitudinal velocity can
deviate from the scaling behavior, $v_{z,{\rm sc}}=z\slash t$, and
with our parametrization, Eq.~(\ref{vap}), fixing $t_0$ is equivalent
to fixing $y_{0}$, the rapidity at $z_0$, as is done also in Refs.
\cite{Venugopalan94,Bolz92}.

\vspace*{0.2in}

The values of $z_0$ and $t_0$ are chosen to fit the pion spectra.  In
Fig.~\ref{ini}, we show the initial profiles for the calculation with
the EOS~A. The main features are the maximum of the energy density at
finite $z$ instead of at $z=0$, and a smooth behavior instead of a
discontinuous one at the edge of matter.  During the calculation, the
velocity of the first vacuum cell is set equal to the velocity of the
nearest cell containing matter.

\vspace*{0.2in}

The profiles of the energy density for the other EOSs look similar.
The parameters are listed in Table \ref{tab2}.  In calculating the
total energy and the baryon number of the initial matter, only the
volume with $\varepsilon(\vec x) > \varepsilon_f$ ($\varepsilon_f$ is
the freeze-out energy density) is included in the integration.  The
results are equal to the total energy and baryon number fluxes through
the freeze-out surface.

%%%%%%%%%%%%%%%%%%%%%%%%%%%%%%%%%%%%%%%%%%%%%%%%%%%%%%%%%%%%%%%%%%
\section{Freeze-Out}
%%%%%%%%%%%%%%%%%%%%%%%%%%%%%%%%%%%%%%%%%%%%%%%%%%%%%%%%%%%%%%%%%%

We define the freeze-out surface as the space-time hypersurface
$\sigma_{\mu}$ of constant energy density.  The value of
$\varepsilon_f$ is chosen in such a way that the mean value of the
temperature on $\sigma_{\mu}$ is $T_f \simeq 140$ MeV. This value for
the temperature is chosen on the basis of calculations comparing the
various mean free paths of the hadrons in the system with the size of
the fireball \cite{Goity89,RajPrak93,Schnedermann94,Sorge95}.

\vspace*{0.2in}

The hadron spectra are calculated by assuming thermal momentum
distributions and chemical equilibrium on the freeze-out surface.  The
invariant momentum distribution of a hadron $h$ is then given by
\cite{Cooper74}
\begin{equation}\label{cf}
E\:\frac{dN}{d^3 p} = \frac{g_h}{(2\pi)^3} \int\limits_{\sigma}
d\sigma_{\mu} p^{\mu} \: \frac{1}{\exp[(p_{\nu}u^{\nu} - \mu)/T_f] \pm
1} \: ,
\end{equation}
where the temperature $T_f (x)$, the chemical potential $\mu (x)$, and
the fluid flow four--velocity $u^\nu(x)$ are determined on the surface
$\sigma_\mu$ from the hydrodynamic calculation.  After calculating the
momentum distribution for each hadron included in the EOS,
Eq.~(\ref{hadroncomp}), the contributions from unstable resonance
decays are added to the stable hadron spectra.  We use the
approximations and decay kinematics described in
Ref. \cite{Sollfrank91}. Finally, we integrate over the experimental
acceptance in $p_T$ and $y$.

%%%%%%%%%%%%%%%%%%%%%%%%%%%%%%%%%%%%%%%%%%%%%%%%%%%%%%%%%%%%%%%%%%%
\section{Photons}
%%%%%%%%%%%%%%%%%%%%%%%%%%%%%%%%%%%%%%%%%%%%%%%%%%%%%%%%%%%%%%%%%%%

The thermal emission rate for photons can be shown to be directly
proportional to the trace of the retarded photon self energy at finite
temperature \cite{GaKap}. In the QGP, the imaginary parts of the
lowest order contributions to the self energy correspond to tree level
QCD Compton and annihilation diagrams, $q\bar{q}\rightarrow g\gamma$,
$q(\bar{q})g\rightarrow q(\bar{q})\gamma$, respectively. Contributions
from these diagrams alone are infrared divergent. However, it has been
shown by Braaten and Pisarski \cite{BraPi90} that these long range
effects are screened at finite temperature.  The above mentioned
diagrams were calculated, including the Braaten--Pisarski resummation,
in Ref.  \cite{Kapusta91,Baier92} for zero baryon chemical potential.
This result was extended in Ref. \cite{Traxler95} to finite baryon
density.  We use the results of Ref. \cite{Traxler95}, since we have
finite baryon density explicitly in our calculation.  However, the
influence of the term containing the chemical potential is small.  For
two quark flavors, the rate is \cite{Traxler95}
\begin{equation}\label{photonqgprate} E_{\gamma}\frac{dR^{\rm
QGP}}{d^3 p} = \frac{5}{9} \: \frac{\alpha \alpha_s
T^2}{2\pi^2}\left(1+\frac{\mu_q^2}{\pi^2T^2}\right) e^{-E_{\gamma}/T}
\: \ln\frac{0.2317 E_{\gamma}}{\alpha_s T} \:, \end{equation} where
$dR = dN/d^4x$.  We use a temperature dependent running coupling
constant
\begin{equation}
\alpha_s = \frac{6\pi}{(33-2n_f)\ln(T/\Lambda_T)} \: ,
\end{equation}
where we take $\Lambda_T = 40$ MeV \cite{Kapusta89} and, as in the
EOS, $n_f = 2.5$.

\vspace*{0.2in}

For the hadron phase, we use the single photon production rates
calculated in Ref. \cite{Kapusta91}.  These calculations were
performed using a pseudoscalar--vector Lagrangian of the form
\begin{eqnarray}
 {\cal L}=|D_\mu \Phi|^2 -m_\pi^2 |\Phi|^2 -{1\over 4}
\rho_{\mu\nu}\rho^{\mu\nu} +{1\over 2} m_{\rho}^2 \rho_\mu\rho^\mu
-{1\over 4}F_{\mu\nu}F^{\mu\nu} \,,
\end{eqnarray}
where $\Phi$, $A_\mu$ and $\rho_\mu$ are complex pseudoscalar, photon
and vector fields, respectively.  Furthermore, $D_\mu = \partial_\mu
-ie A_\mu -ig_\rho \rho_\mu$ is the covariant derivative, and
$\rho_{\mu\nu}$ and $F_{\mu\nu}$ are the vector meson and
electromagnetic field strength tensors, respectively.  Also, $g_\rho
^2/4\pi =2.9$ as determined from the decay $\rho\rightarrow \pi\pi$.

\vspace*{0.2in}

Parameterizations of the rates contributing to $E_{\gamma} dR^{\rm
HG}/d^3 p$ were given as a function of $T$ and $E_\gamma$ in
Ref. \cite{Nadeau92} for the most important processes.  These include
the two scattering processes $\pi + \pi \rightarrow \rho + \gamma$ and
$\pi + \rho \rightarrow \pi + \gamma$.  The latter was only calculated
for virtual pion exchange.  It was shown in Ref. \cite{Xiong92} that
$\gamma$ production in the $\pi$-$\rho$ channel is dominated by the
$a_1$ decays.  Thus, we include the process $\pi + \rho\rightarrow a_1
\rightarrow \pi + \gamma$ by using the parametrization given in
Ref. \cite{Xiong92}.  The decays $\rho \rightarrow \pi\pi\gamma$ and
$\omega \rightarrow \pi^0 \gamma$, which occur during the lifetime of
the fireball, were also included using the parametrization given in
Ref. \cite{Nadeau92}.

\vspace*{0.2in}

The total single photon spectrum is given by integrating over the
total space-time volume, i.e., over all fluid cells with $\varepsilon
\ge \varepsilon_f$:
\begin{eqnarray}
E\frac{dN^\gamma}{d^3p} = \int d^4 x \, \Big\{&&
w(\varepsilon,\rho_{\rm B}) \; E\frac{dR^{\rm QGP}}{d^3p}(p\cdot u, T,
\mu_{\rm B}) \nonumber\\ && \!\!\!\!\!\!\!\!\!\!  +\left[1 -
w(\varepsilon,\rho_{\rm B})\right] E\frac{dR^{\rm HG}}{d^3p}(p\cdot u,
T) \Big\} \: ,
\label{stint}
\end{eqnarray}
where $w$ is defined in Eq. (\ref{defw}).  One should note that the
rates are functions of $p\cdot u=p^\mu u_\mu$, the energy of the
photons in the rest frame of the emitting fluid element. So far, the
rates for the hadron phase do not contain processes involving baryons.
Hence, there is no dependence on $\mu_{\rm B}$. Possible contributions
from baryons to the thermal photon yields were estimated to be small,
even for Pb + Pb collisions at the CERN-SPS in Ref. \cite{Lichard}.

%%%%%%%%%%%%%%%%%%%%%%%%%%%%%%%%%%%%%%%%%%%%%%%%%%%%%%%%%%%%%%%%%%%
\section{Dileptons}
\label{dilepton}
%%%%%%%%%%%%%%%%%%%%%%%%%%%%%%%%%%%%%%%%%%%%%%%%%%%%%%%%%%%%%%%%%%%

We turn now to a discussion of dilepton production from both the
quark-gluon plasma and the hadron gas.  The dominant process in the
QGP is the reaction $q\bar{q}\rightarrow l^+l^-$, which was computed
in lowest order for finite baryon chemical potential to be
\cite{Cleymans87}: \begin{eqnarray} \frac{EdR^{\rm QGP}}{d M^2 d^3 p}
& = & \frac{5}{9} \: \frac{\alpha^2}{8\pi^2} (1+2m_e^2/M^2)\:
\sqrt{1-4m_e^2/M^2}\: e^{-E/T} \nonumber \\ && \ln\:\frac{\{x_- + \exp
[-(E+\mu)/T]\}[x_+ + \exp(-\mu/T)]}{ \{x_+ + \exp [-(E+\mu)/T]\}[x_- +
\exp(-\mu/T)]} \: , \label{diqgp} \end{eqnarray} where $x_{\pm} =
\exp[-(E\pm |p|)/(2T)]$.

\vspace*{0.2in}

At low invariant masses, reactions at order $O(\alpha^2\alpha_S)$
become important \cite{BrPiYa90}.  However, the mass region where
these corrections are significant is dominated by the Dalitz decays of
the final mesons \cite{AlRu92}.  We have not taken these higher order
contributions into account in our hydrodynamical simulations.

For the hadron gas contribution to the dilepton rate, we use the
results of Ref.~\cite{Gale94}.  As in the calculation of the photon
rates, the authors made use of the most general lowest order
Lagrangian with vector and pseudoscalar mesons.  The coupling of these
fields to the electromagnetic current was computed under the
assumption of Vector Meson Dominance. Prior to these calculations, the
only rates computed were those due to $\pi\pi$ annihilation. The
addition of other vector, pseudoscalar, and axial channels enhances
the total rate by at least an order of magnitude away from the mass
region of the light vector mesons.  In doing so, however, one must not
overcount channels already accounted for in the basic $\pi\pi$
reactions \cite{Lichard94}. Therefore, the vector meson decays were
excluded \cite{Lichard96} from the total rate given in
Ref.~\cite{Gale94}. We have also verified that the rates used in our
calculations are in essential agreement with those calculated recently
in Ref.  \cite{Steele96} using a spectral function approach.
         
\vspace*{0.2in}

We adopt the following procedure in incorporating these rates in our
hydrodynamic simulations.  Our starting point is the parametrization
of the total dilepton production rate $dR^{\rm tot}/dM^2(M,T)$ as a
function of temperature $T$ and invariant mass $M^2 = p^2 = (p_1 +
p_2)^2$, as in Ref.~\cite{Gale94}.  Here, $p$, $p_{1}$, and $p_2$ are
the four--momenta of the pair and the single leptons, respectively.

In order to apply the kinematical cuts, we need the momentum
distribution of the pair. To obtain this, we use the relation
\cite{Srivastava94a}
\begin{equation}\label{dihad}
\frac{dR^{\rm HG}}{dM^2 dy p_T dp_T} = \frac{1}{2 M T {\rm K}_1(M/T)}
\: e^{-E/T} \: \frac{dR^{\rm tot}}{dM^2} (M,T)\: ,
\end{equation}
where $E=p_\mu u^\mu$ is the pair energy and K$_1$ is a modified
Bessel function. This relation is valid for reactions in which the
final state contains only the lepton pair, which gives the dominant
contribution in the higher mass region. However, the relation is not
valid for the decays, $h\to h'e^+e^-$, which result in small mass
pairs.  The region in which Eq.~(\ref{dihad}) is a good approximation
was given to be above 300 MeV in invariant mass in
\cite{Srivastava96}.  We have verified that the approximation
Eq.~(\ref{dihad}) adequately reproduces the rates in
Ref.~\cite{Steele96} above an invariant mass of 400 MeV. Below 400
MeV, differences may exceed a factor of 2 and above in some regions of
the phase space.  However, in this region the spectrum is dominated by
Dalitz decays of mesons {\it after freeze-out}, which are accurately
treated separately.

\vspace*{0.2in}

The rates, Eq. (\ref{diqgp}) and Eq. (\ref{dihad}), are integrated
over the space-time volume of the fireball, as in Eq.~(\ref{stint})
for the photons. Here also, we include baryon contributions from only
the QGP phase.  It remains to be seen whether baryons can contribute
significantly to the dilepton yields (see for example,
Ref. \cite{Cassing95} for initial estimates).

\vspace*{0.2in}

The measured dilepton spectra also contain pairs from decays after
freeze-out.  We consider this as a background contribution.  When
comparing our results with the measurement of the CERES-collaboration
\cite{Agakichiev95}, we calculate this background from our
hydrodynamical simulation, instead of using the background estimated
by the CERES-collaboration \cite{Agakichiev95}.  We combine the
freeze-out momentum distributions of $\pi^0$, $\eta$, $\rho^0$,
$\omega$, $\eta^{\prime}$, and $\phi$ as given by Eq.~(\ref{cf}) with
the distributions from the decays of higher lying resonances.

\vspace*{0.2in}

First, we consider the vector meson decays into electron pairs.  The
hydrodynamic calculation is done assuming a fixed mass and zero width
for the resonance states.  In order to calculate the dilepton spectra,
we have to take into account at least the width of the $\rho^0$ meson.
We assume that the mass distribution of the resonance is of the
Breit-Wigner form:
\begin{equation}
\frac{EdN^V}{dM^2 d^3p} = \frac{\xi}{\pi}\:\frac{\Gamma^2_{\rm
tot}(M)}{ (M^2 -m_V^2)^2 + \Gamma^2_{\rm tot}(M)m_V^2} \:
\frac{EdN^{\rm hydro}}{d^3 p} \: ,
\end{equation}
where $m_V$ is the mean mass of the vector mesons.  The normalization
$\xi$ is determined in such a way that the total yield has the value
calculated at the freeze-out.  We use the following mass dependence of
the total $\rho$ width \cite{Sollfrank91}:
\begin{equation}
\Gamma^{\rho}_{\rm tot} (M) = 3.15 \:\frac{(M^2 -4m^2_\pi)^{3/2}}{ 1 +
39.7(M^2 -4m^2_\pi) } \;.
\end{equation}
with $M$ and $m_\pi$ in units of GeV.  For the other vector mesons
($\omega$, $\phi$), constant widths taken from experiment \cite{PDG94}
are used. This is justified, since the experimental resolution of
CERES is much wider than these widths.  In addition to the cuts
$2m_\pi$ ($2m_K$) for $\rho^0$ ($\phi$) at threshold, we apply a cut
$3m_\pi$ for the $\omega$ mass distribution.  The electro-magnetic
decay width in the Vector Meson Dominance model is proportional to the
pair mass \cite{Cleymans87}. Hence, we use the parametrization
\begin{equation}
\Gamma_{V \rightarrow e^+e^-}(M) = \Gamma^{\rm exp}_{V \rightarrow
 e^+e^-} \: \frac{M}{m_V} \: ,
\end{equation}
with the values of $m_V$ and $\Gamma^{\rm exp}_{V \rightarrow e^+e^-}$
taken from Ref. \cite{PDG94}.  Thus, the electron pair distribution is
given by
\begin{equation}\label{vdec}
\frac{EdN^{\rm pair}}{dM^2 d^3p} = \frac{\Gamma_{V \rightarrow e^+
e^-}(M)}{ \Gamma^V_{\rm tot}(M)}(M) \: \frac{EdN^V}{dM^2 d^3 p}\,.
\end{equation}

\vspace*{0.2in}

Next, we consider the Dalitz-decays of the mesons. We use the
differential decay width \cite{Landsberg85}
\begin{eqnarray}\label{land}
\frac{d\Gamma_{i \rightarrow j e^+ e^-}}{dM^2} &=&
\frac{\alpha\Gamma_{i\rightarrow j \gamma}}{3\pi M^2} \:
(1+2m_e^2/M^2)\: \sqrt{1-4m_e^2/M^2}\nonumber \\ &\times&
\left[\left(1 + \frac{M^2}{m_i^2 - m_j^2}\right)^2 -
\frac{4m_i^2M^2}{(m_i^2 - m_j^2)^2}\right] \left| F(M^2)\right|^2 \: .
\end{eqnarray}
If the particle $j$ is a photon $\gamma$, the normalization to
$\Gamma_{i\rightarrow\gamma\gamma}$ results in an additional factor of
2 on the r.h.s.~of Eq.~(\ref{land}).  For $\pi^0 \rightarrow \gamma
e^+ e^-$, we take a linear approximation for the form factor, $F(M^2)
= 1 + 5.5 \:{\rm GeV}^{-2}\times M^2$\,, from Ref. \cite{Landsberg85}.
For $\eta \rightarrow \gamma e^+ e^-$ and $\omega \rightarrow \pi^0
e^+ e^-$, the dipole approximation, $F(M^2) = (1 -
M^2/\Lambda^2)^{-1}$, with $\Lambda_\eta = 720$ MeV and
$\Lambda_\omega = 650$ MeV, respectively, is used.  For $\eta^{\prime}
\rightarrow \gamma e^+ e^-$, the Vector Meson Dominance form factor

\begin{equation}
|F(M^2)|^2 = \frac{m^4_\rho}{ (M^2-m_\rho^2)^2 + m^2_\rho
             \Gamma^{\rho\,2}_{tot}} \:.
\end{equation}
is used.  The pair distribution is then calculated from
\begin{equation}\label{dalitz}
\frac{EdN^{\rm pair}}{dM^2 d^3p} = \frac{1}{\Gamma^{\rm exp}_{\rm
tot}} \: \frac{d\Gamma_{i \rightarrow j e^+ e^-}}{dM^2} (M) \:
\frac{EdN^{i\rightarrow M j}}{d^3 p} (M) \: ,
\end{equation}
where $\Gamma^{\rm exp}_{\rm tot}$ is taken from Ref. \cite{PDG94} and
$EdN^{i\rightarrow M j}/d^3 p\:$ is an isotropic electron pair
distribution resulting from the decay of an unpolarized meson $i$ into
a pair of mass $M$ and a particle $j$, as given in
Ref. \cite{Sollfrank91}.  The incorporation of the experimental cuts
and resolution is described in the appendix.

%%%%%%%%%%%%%%%%%%%%%%%%%%%%%%%%%%%%%%%%%%%%%%%%%%%%%%%%%%%%%%%%%%%
\section{Results}
\label{results}
%%%%%%%%%%%%%%%%%%%%%%%%%%%%%%%%%%%%%%%%%%%%%%%%%%%%%%%%%%%%%%%%%%%

In this section, we discuss the results of our attempt to describe
consistently the hadron and electromagnetic spectra in S+Au collisions
using the hydrodynamic approach.  We fix the parameters for the
initial conditions from the hadron spectra.  In general, good fits are
obtained only for the pion distributions --- the fits to heavier
hadron spectra are significantly worse.  These are discussed in detail
below.  We apply the temperature and velocity distributions
corresponding to these fits to predict the photon and dilepton spectra
for each EOS.  Our results may be summarized as follows.  The existing
photon data are not sufficiently precise to exclude any EOS except the
EOS for an ideal, massless pion gas.  Hopefully, future experiments
will allow for better discrimination. With regard to the CERES
dilepton data, none of the EOSs considered, in conjunction with the
standard leading order dilepton rates, succeed in reproducing the
observed excess of dileptons below the $\rho$ peak.

%%%%%%%%%%%%%%%%%%%%%%%%%%%%%%%%%%%%%%%%%%%%%
\subsection{Hadron Spectra}
%%%%%%%%%%%%%%%%%%%%%%%%%%%%%%%%%%%%%%%%%%%%%

We begin our discussion with the results for the hadron spectra in
S+Au collisions.  In Fig.~\ref{hadrony}, we show the rapidity spectra
for several hadrons obtained from calculations using three different
EOSs.  In choosing the parameters in the calculation, our main
emphasis has been to reproduce the negative particle spectrum. In the
case of the ideal pion gas EOS I, we assume that the negatives consist
of negative massless pions only.  For the EOSs A and B, both of which
exhibit a phase transition, the initial conditions are quite similar,
but for the two purely hadronic EOSs H and I, they differ considerably
(c.f.~Table~\ref{tab2}).

\vspace*{0.2in}

Let us first note the features that are common to the three different
EOSs.  For the negatively charged particles, which are mostly pions,
the calculations are in good agreement with the data from NA35
\cite{Gazdzicki95}.  This tells us that, by a suitable choice of the
initial conditions, the flow of energy density across the freeze-out
surface can be reproduced reasonably well with all the EOSs
considered.  The three pion states carry roughly half of the total
energy.  The other half is carried mainly by the baryons.

\vspace*{0.2in}

The experimental net proton $p-\bar{p}$ distribution in
Fig.~\ref{hadrony} shows a large asymmetry with respect to the center
of momentum.  (The two data points with $y<3$ are identified proton
data from a different experiment NA44~\cite{Murray95} with S + Pb. It
is instructive to view these points together with the net proton
distribution from NA35 \cite{Rohrich94}, because in this region the
anti-proton yield is very small. While NA35 has practically a full
$p_T$-acceptance up to 2 GeV, the NA44 data were extrapolated to the
full $p_T$ region.)  There is a large difference in the target and
projectile size, and we are presently unable to take this into account
in our code with longitudinal mirror symmetry. However, we do not want
to neglect the baryons totally, since they carry a significant part of
the total energy; their influence is felt through the EOS during the
space-time evolution. We reproduce the right total number of baryons
in our calculation.  Also, the baryon rapidity density in the central
rapidity region ($y_{\rm cm}=2.6$) is roughly reproduced.  This should
ensure a reasonable description of the expansion in the central
region, where our calculations of the electromagnetic yields are
compared with measurements.

\vspace*{0.2in}

The $K_s^0$ distribution comes out too large in comparison with
experiment.  This may be attributed to the fact that we have assumed
full chemical equilibrium.  A detailed chemical analysis of strange
particle spectra in the S + Au collisions in Ref. \cite{Letessier95}
shows a strangeness suppression of 0.7 relative to full chemical
equilibrium.  Multiplying the $K_s^0$-spectra and the other strange
particle spectra by this factor would improve the fit.  The abundance
of $K_{\rm s}^0$ mesons is also slightly influenced by the baryon
distributions through the requirement of strangeness neutrality.  The
$K^+$ and $K^0$ have to compensate the net negative strangeness of the
$\Lambda$ baryons.  In the region $3 < y <4$, the overestimation of
the $\Lambda$'s leads to a small additional contribution to the
$K_s^0$'s.

\vspace*{0.2in}

The $\Lambda$ and $\bar{\Lambda}$ rapidity distributions are
influenced both by the strangeness and the baryon number chemical
potentials.  From the rapidity distribution of the protons, we see
that the baryon number, or equivalently the baryon chemical potential,
is too large in the forward rapidity region, because of the
longitudinal symmetry in our calculation.  The excess in the $\Lambda$
spectrum in the region $3<y<4$ is similar to that for protons.  In
order to answer the question of how well strangeness chemical
equilibrium holds for the strange baryons, we should first reduce
$\mu_{\rm B}$ to reproduce the proton distribution in this rapidity
range.  It seems that this would lead to a reasonable agreement for
the $\Lambda$'s, assuming that $\mu_{\rm s}$ would be unchanged.
However, a reduction in $\mu_B$ would not be sufficient to increase
the $\bar{\Lambda}$ yield to the experimental value at rapidities
around 3.

\vspace*{0.2in}

We may summarize our analysis of the strange hadrons by stating that,
for the freeze-out parameters that fit the negative particle yields,
there is agreement in the yield of $\Lambda$'s relative to that of
protons, but the $\bar{\Lambda}$'s are underestimated and the $K_0$'s
are slightly overestimated.  An analysis of S+S collisions at the same
freeze-out temperature, $T_f = 140$ MeV, has led to the same
conclusion \cite{Bolz92,Huovinen96}.

\vspace*{0.2in}

The particle yields are determined in our model by the assumption of
chemical equilibrium until freeze-out.  This is poorly justified in
the late dilute phase, since the strangeness changing cross sections
are small.  Therefore, a study of the chemical behavior in the late
phase deserves further investigation on its own right, but is outside
the scope of this investigation.  However, looking at the pion
distributions, we think that the bulk behavior of the longitudinal
expansion is well reproduced by our model.

\vspace*{0.2in}

The transverse momentum distributions are shown in
Fig.~\ref{hadronpt}.  The overall agreement is quite good with the
exception of the EOS~I. (We will return to discuss this case later.)
The $m_T$-spectra of $\pi^0$s are well fit up to 2 GeV.  For larger
transverse momenta, the three EOSs can also be distinguished.  If one
uses the same initial conditions, then a stiffer EOS would produce
more transverse flow.  However, to fit the longitudinal spectra, we
enhanced the initial energy density for the EOS~B and reduced it for
the EOS~H in comparison to that for the EOS~A (c.f.  Table
\ref{tab2}).  The initial temperatures are roughly the same in each
case, and lead to the similar final $p_T$-slopes.

\vspace*{0.2in}

Although the best agreement is achieved using the EOS~H, fits for the
other EOSs can be improved by fine-tuning the initial conditions.
Further, it is doubtful that particles with very large $p_T$ follow a
hydrodynamical behavior.  Instead, they might originate from semi-hard
processes.  We therefore conclude that, with the exception of the
EOS~I, we cannot rule out any of the remaining three EOSs on the basis
of fits to the $p_T$-spectra.

\vspace*{0.2in}

The ideal pion gas EOS~I produces too flat a slope. The initial
conditions in this case correspond to the same total energy as in the
other three cases (see Table \ref{tab2}).  This was necessary in order
to get the rapidity spectra correct in the central rapidity region.
Since there are no baryons to share the energy, the surplus energy is
converted to transverse kinetic energy.  In short, for the EOS~I, we
could not find initial conditions which reproduce both the rapidity
and $p_T$-spectra simultaneously (see also
Ref. \cite{Dumitru95}). Therefore, the ideal pion gas EOS~I can be
ruled out, on the basis of hadronic data, as being too stiff.

\vspace*{0.2in}

The absolute yields of the other particles in Fig.~\ref{hadronpt}
depend on the assumption of chemical equilibrium, the inadequacy of
which we have discussed above.  This is also corroborated by the the
anti-proton yield, which is also underestimated, like that of the
$\bar{\Lambda}$'s.  We see, however, that the slopes for the different
particles are reasonably reproduced by all three EOSs, supporting the
picture of collective transverse flow present at SPS energies
\cite{Schnedermann92}.

\vspace*{0.2in}

At low $p_T$, especially for the $\Lambda$'s, discrepancies between
calculations and data persist.  The calculation is expected to
overshoot the data on the basis of the excess seen in the rapidity
spectrum in the $3<y<4$ region.  The larger relative weight from this
region in the calculation as compared to the data might also be a
reason for the difference in the slopes.

\vspace*{0.2in}

The experimental $\eta/\pi^0$-ratio for central events
\cite{Lebedev94} has large error bars.  It is reproduced reasonably
well at large $p_T$, but our calculation underestimates it at low
$p_T$. In our model, the $4\pi$ integrated $\eta/\pi^0$-ratio depends
mainly on the freeze-out temperature.  In Table \ref{tab3}, we compare
the total multiplicities with $p+p$ data on meson production
\cite{Aguilar-Benitez91}.  Our $\eta/\pi^0$-ratio is similar to that
in proton collisions and the minimum bias data in S + Au collisions
\cite{Albrecht95a}.

\vspace*{0.2in}

To summarize the discussion of the hadron spectra, we see that with
the exception of the ideal pion gas EOS, the initial conditions can be
adjusted to reproduce the gross behavior of the hadron (negatives)
spectra for a wide class of EOSs.  Details of the spectra depend on
the assumption of chemical equilibrium, which turns out to be poorly
justified at freeze-out temperatures of $T_f\sim 140$ MeV. Chemical
equilibrium between particles and resonances which have a large cross
section may still hold.

%%%%%%%%%%%%%%%%%%%%%%%%%%%%%%%%%%%%%%%%%%%%%
\subsection{Photon Spectra}
%%%%%%%%%%%%%%%%%%%%%%%%%%%%%%%%%%%%%%%%%%%%%

The fundamental difference between the hadron spectra and the photon
spectra is the fact that photons are emitted from collisions of
charged particles during the entire expansion stage. We have
calculated the photon $p_T$-spectra using {\it the same simulation}
from which the hadron spectra were obtained.  We integrate the photons
over the four-dimensional space-time volume, Eq. (\ref{stint}), which
is bounded by the three-dimensional freeze-out surface used for the
calculation of the hadron spectra. The calculations are compared with
the upper bound of the WA80 collaboration on the direct photon
spectrum \cite{Albrecht95b}.  Experimentally, the direct photon yield
is obtained by subtracting photons from the decays of mesons and
baryons.

\vspace*{0.2in}

Since the photon yield depends on the properties of the system as it
expands, we first discuss how the space-time behavior is affected by
the different EOSs.  In Fig.~\ref{energycontour}, contours of constant
energy density in the $zt$-plane at $r=0$ are shown, for all four
EOSs.  The boundaries of the mixed phase are indicated as solid lines.
The freeze-out times at the center $z=0$ are $t_f= 9.4$, 7.9, 7.9, and
7.5 fm/$c$ for EOSs A, B, H, and I, respectively.  At $r=0$,
transverse expansion is absent essentially up to $t_f$.  Similar
freeze-out times for different EOSs indicate that the longitudinal
expansion in the central region is dominated by the initial velocity
gradient.  At large $z$, the QGP equations of state produce a
long-living tail. This is the result of a weaker longitudinal
acceleration, due to the smaller pressure gradient in the mixed phase
as compared with that in a calculation employing the hadron gas EOS
without phase transition.

\vspace*{0.2in}

The space-time volume of the mixed phase in the EOS~B is somewhat
larger than that for the EOS~A, despite the fact that it has a smaller
latent heat, which allows, per unit volume, a faster conversion of the
mixed phase into the hadron gas. However, for the EOS~B, the mixed
phase is reached at a later time and with so much larger volume that
the conversion takes more time and produces a bigger final volume than
that for the EOS~A.

\vspace*{0.2in}

In Fig.~\ref{photon}a, we show the direct photon spectra for the
EOS~A. From the individual contributions from the different phases, it
is clear that hadronic processes dominate the production of single
photons in this case.  In Ref. \cite{Kapusta91}, it was shown that at
the same temperature, emission rates per unit volume are roughly the
same in a QGP and a hadron gas.  Inclusion of the $a_1$ mesons
enhances photon production for energies $E_\gamma > 1$ GeV
\cite{Xiong92}, leading, at $T_c$, to a considerably larger photon
production rate in the hadron gas than in the QGP.  Also, the high
temperature phase above $T_c$ lasts for a short time in this case, and
the produced matter spends most of its lifetime in the hadron gas
phase.  These two features lead to the dominance of photon emission
from the hadronic phase.  The higher temperature of the initial QGP
phase shows up as a flatter $p_T$ slope in the plasma contribution,
but even at the largest $p_T$ values, it is clearly below the hadron
gas contribution.  The total yield is in agreement with the upper
bound provided by the data.

\vspace*{0.2in}

In Fig.~\ref{photon}b, the individual contributions are shown for the
EOS~B. In this case, the hadrons from the mixed phase dominate photon
production.  The QGP yields are similar to those with the EOS~A. At
large $p_T$, the QGP contribution is as large as the hadron
contribution from the mixed phase.  There is contribution from the
pure hadron gas phase, because the transition from the mixed phase to
the hadron phase and freeze-out takes place simultaneously at the same
temperature.

\vspace*{0.2in}

Figure \ref{photontot} shows the total single photon spectra for the
four different EOSs.  The situation is qualitatively similar to that
for the $p_T$-spectrum of $\pi^0$'s in Fig. \ref{hadronpt}.  However,
quantitative differences exist. A close inspection reveals that not
only the results for the EOS~I, but also for the other three EOSs,
differ from one another.  The total yield with the EOS~B lies a factor
2--3 below that of the EOS~A, whereas the EOS~H produces 2--10 times
more photons depending on the $p_T$-region. The yield for the EOS~I
lies orders of magnitude above.

\vspace*{0.2in}

In principle, the dependence of the photon production on $T_c$ could
be used to determine $T_c$ from the data.  However, one can see from
Fig.~\ref{photontot} that the present upper bound on the single photon
yield rules out only the ideal pion gas EOS~I. (Recall that this EOS
is also ruled out by the transverse momentum spectrum of negative
hadrons.)  From our results of the photon spectra, we cannot claim
evidence for a phase transition, in contrast to the claims in
Refs. \cite{Dumitru95,Srivastava94b,Shuryak94}.  The main point here
is that, if a reasonable amount of degrees of freedom are taken into
account in a hadron gas, the increase of temperature with energy
density is reduced.  This is clearly seen by comparing results for the
EOS~H with those for the EOS~I.  The large photon yield in the case of
the EOS~I is due to the very large initial temperature, $T_i = 400$
MeV, as compared to $T_i = 250$ MeV for the EOS~H.

\vspace*{0.2in}

In the extreme case of the Hagedorn bootstrap model \cite{Hagedorn94},
we have a limiting temperature $\sim 150$ MeV in a hadron gas, with
arbitrarily high energy densities.  For limited energy densities, such
an EOS would lead to a small photon production.  Therefore, the
results do not readily attest to the existence or the absence of a
phase transition.  With the present precision, one can only rule out
high initial temperatures.  However, since the rates from the hadron
gas and the QGP differ, the correlation between the total yield and
the slope of the transverse momentum distribution will differ for the
purely hadronic EOS and an EOS with a phase transition. For this
reason, improving the experimental upper limit and measuring the yield
of direct photons is very important.

\vspace*{0.2in}

In discussing the dependence of the photon production on the
transition temperature $T_c$, one should keep in mind not only the
limitations of hydrodynamics, but also the uncertainties in the rate
calculations both in the QGP and a hadron gas near $T_c$.  Our
discussion above is based on the considerable difference in the rates
at the same temperature between perturbative QCD results
\cite{Traxler95} and leading order estimates in a hadron gas
\cite{Kapusta91,Xiong92}.

\subsubsection{Comparison with other works}

Finally, we want to compare our results with other calculations of
photon yields in the S + Au collisions, which have used hydrodynamics
to describe the evolution of the produced matter.  These earlier
calculations were compared with the preliminary data of WA80
\cite{Santo94}.  The new analysis by WA80 gives upper bounds, which
are compatible with the preliminary data.

\vspace*{0.2in}

First, we confirm the result of Ref. \cite{Shuryak94} that the
absolute normalization of the photon yield is sensitive to $T_c$ and
to the EOS. The calculation in Ref. \cite{Shuryak94} was done using a
one-dimensional Bjorken--like hydrodynamical scenario, whereas we
employ a three-dimensional expansion.  The small differences in the
photon yield from our results can be attributed to the differences in
the hydrodynamical solutions.  In Ref.  \cite{Shuryak94}, consistency
with the preliminary data of WA80 was achieved with $T_c \approx 200$
MeV, while lower critical temperatures of around 150 MeV led to an
underprediction of the preliminary WA80 data.  Therefore, the
possibility of a long-lived mixed phase, of duration 30-40 fm/$c$, was
suggested there.  For the EOSs used in our three-dimensional
calculations, $T_c$ lies in the range 140-165 MeV and the duration of
the mixed phase does not exceed 10 fm/$c$. Since the present data
provides an upper bound only, a long-lived scenario is not necessary
with the lower values of $T_c$ which we have been using.

\vspace*{0.2in}

Our results roughly agree with the investigation in
Ref. \cite{Arbex95} as well.  However, Arbex et al. \cite{Arbex95}
only investigated an EOS with $T_c$ of 200 MeV. Since, as noted above,
the absolute normalization of the photon spectra is sensitive to the
critical temperature, their results agreed with the preliminary data
of WA80.

\vspace*{0.2in}

In Refs. \cite{Dumitru95} and \cite{Srivastava94b}, an agreement with
the preliminary data of WA80 was achieved using an EOS with $T_c$
around 160 MeV, while our calculation with EOS~A would underpredict
this preliminary data by roughly a factor of 4. In
Ref. \cite{Srivastava94b}, a rather low freeze-out temperature, $T_f =
100$ MeV, was chosen.  We have checked that the photon yield increases
by a factor of $\sim 1.5$ when the calculation is extended from a
freeze-out temperature $T_f=140$ MeV to $T_f=100$ MeV. In Ref.
\cite{Srivastava94b}, the neglect of baryons and the smaller number of
mesons in the EOS leads to considerably longer lifetimes for the mixed
and hadron phases.  In addition, our three-dimensional calculations
lead to more rapid cooling than that obtained using the scaling
expansion, even when transverse expansion is included.  A combination
of these effects can explain the larger yield in
Ref.~\cite{Srivastava94b}.

%%%%%%%%%%%%%%%%%%%%%%%%%%%%%%%%%%%%%%%%%%%%%
\subsection{Dielectrons}
%%%%%%%%%%%%%%%%%%%%%%%%%%%%%%%%%%%%%%%%%%%%%

The other electromagnetic signal, measured in the S+Au collisions by
the CERES collaboration \cite{Agakichiev95}, is the dielectron mass
distribution.  In these measurements, the dilepton background from the
decays of final mesons is not subtracted, because the amounts of
different mesons are not precisely known.  Thus, we have two parts in
the final dilepton yield.  First, the emission during the lifetime of
the fireball and, second, the electromagnetic decays of hadrons after
the decoupling.  The latter contribution is shown in
Fig.~\ref{ceresback} for the EOS~A. This background is similar for the
other EOSs, since the final hadron spectra are reproduced in each case
by tuning the initial conditions, as discussed above.  The calculation
of the decay dilepton spectrum is described in Sec. VII and the
appendix. All hadrons which produce lepton pairs are considered.  Both
the thermal contribution and the contributions from the decays from
heavier hadrons are included.

\vspace*{0.2in}

Our calculated background in Fig.~\ref{ceresback} is basically in
agreement with the estimated background in
Ref. \cite{Agakichiev95}. However, there are some differences in the
total yields of the decaying mesons, since in our calculation the
meson multiplicities are given by the calculated freeze-out
conditions, mainly the assumed temperature, the effect of the baryon
chemical potential being small. On the other hand, the CERES
collaboration used the meson-to-$\pi^0$ ratios from p+p collisions
\cite{Aguilar-Benitez91} to fix the multiplicities of mesons from the
measured $\pi^0$ spectrum \cite{Santo94}.  In Table \ref{tab3}, we
present the meson yields normalized to the $\pi^0$ yield.  Our $\eta$
multiplicity is somewhat higher than in $p+p$ collisions, while that
of the $\rho^0$ is slightly smaller.  For the $\omega$ meson, the
ratio is 3/4. The main differences in the input yields are those for
the $\eta^{\prime}$ and $\phi$. The contributions from the
$\eta^{\prime}$ are negligible compared to those from the other
dilepton sources. Thus, the only important difference with the CERES
background is a factor of 4 in the $\phi$--mass region; however, this
is still in agreement with the data.

\vspace*{0.2in}

In Fig.~\ref{dielectron}a, we show the dileptons radiated during the
lifetime of the fireball.  These results are folded with the CERES
cuts and the CERES resolution. Here, we see the same systematics as in
the case of photons; the hadronic contribution dominates the yield.
The largest contribution comes from the $\pi$--$\pi$ annihilation to
dielectrons via the $\rho$--form factor (Vector Dominance); this is
seen as a peak at the $\rho$ mass. As for the photons, the
contribution from the pure hadron gas and the hadronic contribution in
the mixed phase are equally important.

\vspace*{0.2in}

The sum of the background and the thermal emission is shown in
Fig.~\ref{dielectron}b.  We first note that thermal emission roughly
fills the gap between the background and the data around and above the
$\rho$ mass, even though the calculated results tend to lie at the
lower bound of the errors, especially for the EOS~B.  Note that the
systematic and the statistical errors are shown separately
\cite{Tserruya}.

\vspace*{0.2in}

In the mass region between 200 and 600 MeV, our calculated results lie
clearly below the data.  While the calculation has a dip in this
region between the contributions from the Dalitz pairs and the vector
mesons, the data is flat and smooth.  There have been several
suggestions for the origin of the excess over the expected sources in
this invariant mass region
\cite{Cassing95,Koch93,Li95,Kapusta95,Redlich95}.  An interesting
possibility for the explanation is a shift of the vector meson masses
associated with the expected restoration of chiral symmetry as the
transition temperature $T_c$ is approached in the hadron gas.  We have
not tried to include this in our calculation.  With an appropriate
parametrization of the temperature and/or density dependence of the
hadron masses, we most likely would be able to reproduce the data, but
a consistent treatment in a hydrodynamical approach would require the
incorporation of density dependent masses to also calculate the EOS
and the decay rates.  This is beyond the scope of the present work,
and its proper implementation requires a major effort, which will be
taken up separately.

\vspace*{0.2in}

Our conclusions for the EOS from the dilepton calculations are similar
to those for photons.  The absolute yields are sensitive to $T_c$, and
the results for the EOS~B with $T_c=140$ MeV are below the data at all
mass values.  However, the differences are not as pronounced as for
the photons, since the contributions from the decays of final hadrons
have not been subtracted.

\subsubsection{Comparison with other works}

The excess production of low-mass dilepton pairs in S + Au collisions
was recently addressed in Ref. \cite{Srivastava96} using a
one-dimensional Bjorken-like expansion.  In the invariant mass range
0.2 GeV $< M <$ 0.6 GeV, where the excess over expected sources is
most evident, the data were underpredicted by about a factor of 4.
Due to the high initial temperature ($T_i=380$ MeV) required to obtain
similar results in a purely hadronic scenario, the case in which a QGP
is admitted ($T_i=198.7$ MeV, $T_c=160$ MeV, $T_f=140$ MeV, and a
mixed phase duration $\sim 13$ fm/$c$) was favored in this work.

Where the deviations from the data are largest, our three-dimensional
calculations (with the same standard rates as above) underpredict the
data by a much larger factor (of about 8-10).  Several sources for the
differences from Ref. \cite{Srivastava96} may be cited. These include
the use of a more realistic EOS, a shorter duration of the mixed
phase, and features specific to the three-dimensional flow of the
matter emitting the lepton pairs.  Clearly, a combination of these
effects results in dilepton yields that are lower than in the case of
a one-dimensional Bjorken evolution, even in the case where an EOS
admitting a phase transition to the QGP is used. We also calculate the
background contributions from the calculated hadron spectra, whereas
the CERES background is used in Ref.~\cite{Srivastava96}.

A comparison of our hydrodynamical results with the alternative
sequential scattering models (also termed as cascade or transport
models) depends on the extent to which thermalization is achieved in
the latter approach. Specific medium modifications of the vector meson
properties, in particular a decrease in their masses, have been found
to yield a satisfactory description of the CERES data
\cite{Li95,Cassing95}. Whether a similar approach can be
satisfactorily adopted in hydrodynamical simulations is a challenging
future task.

%%%%%%%%%%%%%%%%%%%%%%%%%%%%%%%%%%%%%%%%%%%%%%%%%%%%%%%%%%%%%%%%%%%%
\section{Conclusion}
%%%%%%%%%%%%%%%%%%%%%%%%%%%%%%%%%%%%%%%%%%%%%%%%%%%%%%%%%%%%%%%%%%%%

The aim of this work was to establish the extent to which one can
constrain the EOS from the experimental data for S + Au collisions
from a simultaneous description of the hadron and electromagnetic
spectra using hydrodynamics.  We have shown that, in general, the
influence of the EOS on hadronic spectra can be counterbalanced by
choosing different initial conditions.  A simultaneous calculation of
the electromagnetic signals can, in principle, distinguish between the
different EOSs.  However, the present experimental resolution allows
us to rule out only extreme cases, such as the ideal pion gas EOS with
only a few degrees of freedom.  Also, the dilepton yield for the QGP
equation of state with $T_c\simeq 140$ MeV, the EOS~B, tends to fall
below the data in the vector meson mass region, indicating an
effective lower bound of 140 MeV for $T_c$, if the transition exists.

\vspace*{0.2in}

The constraint that can be drawn from the single photon data is that
the initial temperature cannot be too high.  The present data rules
out temperatures above 250 MeV. This limit on the initial temperature
can be achieved only if a large number of degrees of freedom is
involved, be it in the form of quarks and gluons, or in the form of a
large enough number of hadrons.  However, if the data can be improved,
the two cases can be distinguished, since the total emission from a
hadron gas is larger than that from the QGP.  In the total yield, the
difference between a pure hadronic EOS and an EOS with a phase
transition increases with decreasing $T_c$.

\vspace*{0.2in}

The behavior of the dilepton spectrum in the mass region between 200
and 600 MeV shows that the description of the hot and dense strongly
interacting matter near $T_c$ in terms of the free-space parameters is
not adequate.  With our present hydrodynamical approach, the dilepton
spectrum can be explained only in the mass region of the vector
mesons.  The large experimental dilepton yield below the $\rho$ mass
may indicate medium modifications of the particle properties.  These
effects can be included in hydrodynamical calculations, but for a
consistent calculation, the EOS must be modified accordingly.

\section*{ACKNOWLEDGMENT}

We gratefully acknowledge helpful discussions with G.E. Brown,
C. Gale, P.  Lichard, H. Sorge, J. Steele, and I. Zahed.  We thank
E.K. Popenoe for a careful reading of the paper.  J.S. acknowledges
the kind hospitality of the INT in Seattle and the Department of
Physics at SUNY in Stony Brook, where part of this work was done.
P.H~'s work is supported by a grant from the Academy of Finland.  The
research of M.P.~ is supported by the U.S. Department of Energy under
Grant number DE-FG02-88ER-40388.  R.V.~'s research at the INT is
supported by DOE Grant DE--FG06--90ER40561.

\newpage

\section*{APPENDIX}

%%%%%%%%%%%%%%%%%%%%%%%%%%%%%%%%%%%%%%%%%%%%%%%%%%%%%%%%%%%%%%%%%%%%
\section*{Cuts and Resolution for Comparison with CERES\label{cuts}}
%%%%%%%%%%%%%%%%%%%%%%%%%%%%%%%%%%%%%%%%%%%%%%%%%%%%%%%%%%%%%%%%%%%%

The kinematic cuts of the CERES experiment are cuts in the momenta of
the electron and the positron \cite{Agakichiev95}. We incorporate them
in the following way. We take $dN/(dM^2 dy p_T dp_T)$ of the pair
calculated either from the background contribution (see Eq.~\ref{vdec}
and Eq.~\ref{dalitz}) or that resulting from the fireball during its
lifetime (see Eq.~(\ref{diqgp}) and Eq.~(\ref{dihad})).  Going to the
pair rest frame, we assume an isotropic momentum distribution.  Thus,
the single electron momentum distribution $dN/(dM^2 dy_1 p_{1T}
dp_{1T})$, when the pair mass is $M$, is given by
\begin{eqnarray}
\frac{dN}{dM^2 dy_1 p_{1T} dp_{1T}} & \! = \! &
\frac{M}{2\pi\sqrt{(M^2-4m_e^2)}}\:\int\limits_{y^-}^{y^+} dy
\int\limits_{(m^-_T)^2}^{(m^+_T)^2} dm_T^2 \:\:\:
\theta(p_{2T} - p_T^{\rm cut}) \nonumber \\
&\times&
\theta(\eta^{\rm cut}_{\rm max} - y_2 - y_{\rm cm}) \:
\theta( y_2 + y_{\rm cm} - \eta^{\rm cut}_{\rm min}) \:
\theta( \vartheta_{12}^{\rm lab} - \Theta^{\rm cut})
\nonumber \\ [0.9ex]
&\times&\frac{1}{
\sqrt{p_T^2p_{1T}^2 -[M^2/2- m_T m_{1T} \cosh(y_1 -y)]^2}}
\:\frac{dN}{dM^2 dy p_T dp_T} \,,
\end{eqnarray}
where $y$, $y_i$ are the rapidities in the fireball rest frame.
Further, 
\begin{eqnarray}
y^{\pm} &=& y_1 \pm \sinh^{-1}[\sqrt{(M^2 - 4m_e^2)}/(2m_{1T})]
\nonumber \\
m^{\pm}_T &=& \frac{M}{2} \: \frac{M m_{1T}\cosh(y-y_1) \pm
p_{1T}\sqrt{M^2 - 4m^2_e - 4m_{1T}^2\sinh^2(y - y_1)}}{
m_{1T}^2\sinh^2(y - y_1) + m^2_e}  \,.
\end{eqnarray}
The opening angle of the electrons in the laboratory system,
$\vartheta_{12}^{\rm lab}$, neglecting the
electron mass, is given by
\begin{equation}
\cos(\vartheta_{12}^{\rm lab}) =
1 - \frac{M^2}{2|p_1|^{\rm lab}|p_2|^{\rm lab}}
\end{equation}
with 
$|p_i|^{\rm lab} = p_{iT} \sqrt{1 + \sinh^2(y_i + y_{\rm cm})}$.

\vspace*{0.2in}

The final spectrum of the pairs in the approximation
$\eta_{e^\pm}=y_{e^\pm}$  is then 
\begin{equation}
\frac{dN^{\rm cut}}{dM d\eta} = \frac{2M}{\Delta\eta^{\rm exp}}
\int\limits_{p_T^{\rm cut}}
dp_{1T} \: p_{1T} \int\limits_{\eta^{\rm cut}} dy_1 \:
\frac{dN}{dM^2 dy_1 p_{1T} dp_{1T}}   \,.
\end{equation}

\vspace*{0.2in}

In order to compare with the CERES experiment, we finally have to 
fold the calculated results with the detector resolution. We use a 
Gaussian folding
\begin{equation}
\frac{dN^{\rm CERES}}{dM d\eta} (M) =
\int dM' \frac{1}{\sqrt{2\pi} \sigma(M')} \: \exp\left(
- \frac{(M-M')^2}{2\sigma^2(M')}\right) \:
\frac{dN^{\rm cut}}{dM d\eta} (M') \:,
\end{equation}
with the width $\sigma(M)$ taken to be the
detector resolution kindly provided
to us by the CERES collaboration \cite{Tserruya}.

\newcommand{\IJMPA}[3]{{\it Int.~J.~Mod.~Phys.} {\bf A#1}, (#2) #3}
\newcommand{\JPG}[3]{{\it J.~Phys.} {\bf G#1}, (#2) #3}
\newcommand{\AP}[3]{{\it Ann.~Phys.(NY)} {\bf {#1}}, (#2) #3}
\newcommand{\NPA}[3]{{\it Nucl.~Phys.} {\bf A{#1}}, (#2) #3}
\newcommand{\NPB}[3]{{\it Nucl.~Phys.} {\bf B{#1}}, (#2) #3}
\newcommand{\PLB}[3]{{\it Phys.~Lett.} {\bf B{#1}}, (#2) #3}
\newcommand{\PRv}[3]{{\it Phys.~Rev.} {\bf {#1}}, (#2) #3}
\newcommand{\PRC}[3]{{\it Phys.~Rev.} {\bf C{#1}}, (#2) #3}
\newcommand{\PRD}[3]{{\it Phys.~Rev.} {\bf D{#1}}, (#2) #3}
\newcommand{\PRL}[3]{{\it Phys.~Rev.~Lett.} {\bf {#1}}, (#2) #3}
\newcommand{\PR}[3]{{\it Phys.~Rep.} {\bf {#1}}, (#2) #3}
\newcommand{\ZPC}[3]{{\it Z.~Phys.} {\bf C{#1}}, (#2) #3}
\newcommand{\ZPA}[3]{{\it Z.~Phys.} {\bf A{#1}}, (#2) #3}
\newcommand{\JCP}[3]{{\it J.~Comp.~Phys.} {\bf A{#1}}, (#2) #3}
\newcommand{\HIP}[3]{{\it Heavy Ion Physics} {\bf {#1}}, (#2) #3}

{}
\newpage

%%%%%%%%%%%%%%%%%%%%%%%%%%%%%%%%%%%%%%%%%%%%%%%%%%%%%%%%%%%%%%%%%
%%%%%%%%%%%%%%%%%%%%%%%%%%%%%%%%%%%%%%%%%%%%%%%%%%%%%%%%%%%%%%%%%
\noindent {TABLE \ref{tab1}.~~}
Some physical features characterizing various model EOSs.

\begin{center}
\begin{tabular}{|l|c|c|c|c|}
\hline
           &EOS A        & EOS B   & EOS H   & EOS I\\ \hline
                      &$\pi, K, \eta, \rho, \omega, K^*$
                      &$\pi, K, \eta, \rho, \omega, K^*$
                      &$\pi, K, \eta, \rho, \omega, K^*$
                      & \\
Hadronic              &$p, n, \eta^{\prime}, \phi, \Lambda, \Sigma$
                      &$p, n, \eta^{\prime}, \phi, \Lambda, \Sigma$
                      &$p, n, \eta^{\prime}, \phi, \Lambda, \Sigma$
                      & \\
Degrees of freedom    &$ \Delta, a_1, \Xi, \Sigma(1385)$
                      &$ \Delta, a_1, \Xi, \Sigma(1385)$
                      &$ \Delta, a_1, \Xi, \Sigma(1385)$
                      &$\pi$ \\ \hline
Number of               &       &        &         &       \\
QGP degrees of freedom  & 31    &  31    &  ---    &  ---  \\ \hline
Mean field repulsion      &     &        &         &     \\
$K$ [fm$^3$ MeV]          & 450 & 660  & 450     & 0   \\ \hline
Bag constant              &     &      &         &      \\
$B^{1/4}$ [MeV]           & 235 & 200  & ---     & ---  \\ \hline
$T_c$ [MeV]               & 165 & 140  &$\infty$ &$\infty$\\ \hline
$\mu_{{\rm B}\,c}$ [MeV]  & 1770 & 1290 &$\infty$ &$\infty$\\ \hline
$\Delta \varepsilon$ [GeV/fm$^3$] & 1.4& 0.8 & 0  &0 \\ \hline
\end{tabular}
\end{center}
\newpage

%%%%%%%%%%%%%%%%%%%%%%%%%%%%%%%%%%%%%%%%%%%%%%%%%%%%%%%%%%%%%%%%%%%%

\noindent {TABLE \ref{tab2}.~~}
Variables used for the parametrization of the initial and freeze-out
conditions for the different EOSs. The first part contains the
free parameters, the second part some deduced quantities, and the
third part the freeze-out energy density, with the corresponding
temperature and baryon chemical potential averaged over the
freeze-out hypersurface. The symbols are explained in the text.

\begin{center}
\begin{tabular}{|l|c|c|c|c|}
\hline
                          &EOS A        & EOS B   & EOS H   & EOS I\\ 
\hline
$r_0$ [fm]                & 3.65        & 3.65    & 3.65    &3.65\\
$a_r$ [fm]                & 0.50        & 0.50    & 0.50    &0.50 \\
$z_0$ [fm]                & 0.80        & 0.80    & 1.2     &1.97\\
$a_z$ [fm]                & 0.13        & 0.13    & 0.13    &0.13\\
$\varepsilon_0$[GeV/fm$^3$]& 7.0        & 8.0     & 5.0     &3.3\\
$b_0$ [fm$^{-3}$]         & 1.10        & 1.16    & 0.75    &0\\
$t_0^{-1}$ [c/fm]         & 1.06        & 0.95    & 0.71    &0.41\\ 
\hline
$y(z=z_0)$                & 0.85        & 0.76    & 0.85    &0.80\\
$E_{\rm tot}$ [GeV]       & 914         & 923     & 904     &994\\
$B_{\rm tot}$             & 105         & 104     & 104     & 0\\
$S_{\rm tot}$             & 3880        & 4230    & 3650    &3035\\
$S/B$                     & 37          & 41      & 35      & --\\
$T_i(z=0)$ [MeV]          & 238         & 249     & 248     &400\\ 
\hline
$t_f$ [fm/c]              & 9.4         & 7.9     & 7.9     & 7.5 \\
$\varepsilon_{\rm dec}$[GeV/fm$^3$]&0.165 &0.15   & 0.165   & 0.05\\
$\langle T_f\rangle$ [MeV] & 142        & 141     & 142     & 140\\
$\langle \mu_{{\rm B}\,f}\rangle$ [MeV]&230 &210   & 235    & 0\\ 
\hline
\end{tabular}
\end{center}
\newpage

%%%%%%%%%%%%%%%%%%%%%%%%%%%%%%%%%%%%%%%%%%%%%%%%%%%%%%%%%%%%%%%%%%

\noindent {TABLE \ref{tab3}.~~} Total multiplicities used to estimate
the background to the dielectron spectrum.  The total yields are
normalized to the $\pi^0$ yield.  The EOS A is used in the
calculation.  For comparison, the total production cross section in
$p+p$ collisions at 400 GeV incident momentum are shown in the first
column (the data are taken from NA27 \cite{Aguilar-Benitez91}).
\vspace*{0.05in}

\begin{center}
\begin{tabular}{|c|c|c||c|c|}
\hline
    &\multicolumn{2}{c||}{Hydro}   & \multicolumn{2}{c|}{NA27} \\ 
\hline
    & total  & relative  & total cross & relative \\
    & multiplicity & to $\pi^0$& section [mb] & to $\pi^0$ \\ \hline
$\pi^0$        & 212         & 1.000 & 127.2 & 1.000    \\
$\eta$         & 19.4        & 0.092 & 9.8   & 0.077    \\
$\rho^0$       & 18.6        & 0.088 & 12.6  & 0.099    \\
$\omega$       & 15.8        & 0.074 & 12.8  & 0.101    \\
$\eta^{\prime}$&  2.0        & 0.009 &-- & --           \\
$\phi$         &  4.1        & 0.019 & 0.62  & 0.005    \\ \hline
\end{tabular}
\end{center}

\refstepcounter{table}\label{tab1}
\refstepcounter{table}\label{tab2}
\refstepcounter{table}\label{tab3}

\newpage

%%%%%%%%%%%%%%%%%%%%%%%%%%%%%%%%%%%%%%%%%%%%%%%%%%%%%%%%%%%%%%%%%%

\section*{Figure Captions}

\noindent {Fig. \ref{phaseeos}.~~}
Phase diagrams for two equations of state with a first order phase
transition.
The meaning of the parameters is explained in the text. 
For the EOS A, $K = 450$ MeV fm$^3$, $B^{1/4} = 235$ MeV, 
and for the EOS B, $K = 660$ MeV fm$^3$, $B^{1/4} = 200$ MeV.

\vspace*{0.2in}

\noindent {Fig. \ref{eosfigure}.~~}
Pressure $P$ as a function of baryon $\rho_{\rm B}$ and energy 
densities $\varepsilon$ for the EOS A (a) and the EOS B (b).

\vspace*{0.2in}

\noindent {Fig. \ref{ini}.~~}

Initial conditions for S + Au collisions at $200\ A$ GeV/$c$ for the
case of the EOS A. The different panels show (a) the rapidity $y$, (b)
the longitudinal velocity $v_z$, (c) the energy density $\varepsilon$,
and (d) the baryon density $\rho_{\rm B}$ as a function of the
longitudinal coordinate $z$; (e) the baryon density $\rho_{\rm B}$ as
a function of rapidity $y$; (f) the energy density $\varepsilon$ as a
function of radius $r$.

\vspace*{0.2in}

\noindent {Fig. \ref{hadrony}.~~} Rapidity distributions for several
hadrons compared with a symmetric hydrodynamical calculation. The
equations of state are EOS A (solid line), EOS B (dotted line),
hadronic EOS H (short dashed line), and ideal pion gas EOS I (long
dashed line). The data are measured by the following groups: Negative
charged particle data are from NA35 \cite{Gazdzicki95}, the strange
hadron ($\Lambda$, $\bar{\Lambda}$, $K_s^0$) data from NA35
\cite{Alber94}, the net proton ($p - \bar{p}$) data, with $y > 3$ from
NA35 \cite{Rohrich94}, and the two points of proton data with $y < 3$
are from NA44 \cite{Murray95}.

\vspace*{0.2in}

\noindent {Fig. \ref{hadronpt}.~~} Transverse momentum distributions
for several hadrons compared with a symmetric hydrodynamical
calculation. Curve designations are as in Fig.  \ref{hadrony}. The
data are measured by the following groups: $\pi^0$ data for central S
+ Au collisions from WA80 \cite{Santo94}, the strange hadron
($\Lambda$, $\bar{\Lambda}$, $K_s^0$) data \cite{Alber94}, and the
anti-protons are from NA35 \cite{Guenther95}, and the $\eta/\pi^0$
ratio for central S + Au collisions from WA80 \cite{Lebedev94}.

\vspace*{0.2in}

\noindent {Fig. \ref{energycontour}.~~} Contours of constant energy
density in the $zt$--plane for calculations with the EOS A (a), EOS B
(b), EOS H (c), and EOS I (d).  Contours counted outward correspond to
energy densities of 4.0, 2.0, 0.9, 0.4, 0.15 GeV/fm$^3$ for (a)
through (c), and to 2.0, 0.9, 0.4, 0.15, 0.05 GeV/fm$^3$ for (d). The
solid lines indicate the transition from the QGP to the mixed phase
and from the mixed phase to the hadronic phase. The dashed line
corresponds to freeze-out. In (b), freeze-out occurs at the transition
from the mixed to the hadronic phase. Note the different scale in (b).

\vspace*{0.2in}

\noindent {Fig. \ref{photon}.~~} Single photon $p_T$-spectra compared
with the upper bound of WA80 data \cite{Albrecht95b}; (a) EOS A and
(b) EOS B.  The different contributions shown are: HG hadron gas,
M(HG) hadronic part of the mixed phase, M(QGP) QGP part of the mixed
phase, QGP, and the total spectrum (solid line).

\vspace*{0.2in}

\noindent {Fig. \ref{photontot}.~~}
Total spectrum of single photons for the different EOSs.
The data are as in Fig. \ref{photon}.

\vspace*{0.2in}

\noindent {Fig. \ref{ceresback}.~~}
Estimate of the background to the dielectron spectrum from the meson
decays after freeze-out as calculated from the hydrodynamical result
for the EOS A.  The kinematic cuts and the detector resolution of the
CERES experiment \cite{Agakichiev95} are incorporated.

\vspace*{0.2in}

\noindent {Fig. \ref{dielectron}.~~} Dielectron spectra compared with
the measurement of the CERES collaboration \cite{Agakichiev95}.
Kinematic cuts and detector resolution are incorporated.  (a)
Contributions during the lifetime of the fireball.  (b) Total
dielectron spectrum including our background estimate for the
different equations of state.

\refstepcounter{figure}\label{phaseeos}
\refstepcounter{figure}\label{eosfigure}
\refstepcounter{figure}\label{ini}
\refstepcounter{figure}\label{hadrony}
\refstepcounter{figure}\label{hadronpt}
\refstepcounter{figure}\label{energycontour}
\refstepcounter{figure}\label{photon}
\refstepcounter{figure}\label{photontot}
\refstepcounter{figure}\label{ceresback}
\refstepcounter{figure}\label{dielectron}


\begin{thebibliography}{99}


\bibitem{Landau53}
L.D.~Landau, {\it Izv.~Akad.~Nauk.~SSSR} {\bf 17} (1953) 51.

\bibitem{Stoecker86}
For reviews see:\\
H.~St\"ocker and W.~Greiner, \PR{137}{1986}{277};\\
J.P.~Blaizot and J.Y.~Ollitraut, in \cite{Hwa91}, p.~393.

\bibitem{Clare86}
R.B.~Clare and D.~Strottman, \PR{141}{1986}{177}.

\bibitem{Hwa91}
See for example, {\it Quark-Gluon Plasma} and 
{\it Quark-Gluon Plasma 2},
R.C.~Hwa (ed.), World Scientific, Singapore, 1990, 1995, 
respectively.

\bibitem{Alber94}
T.~Alber et al. (NA35 collaboration), \ZPC{64}{1994}{195}.

\bibitem{Rohrich94}
D.~R\"ohrich et al. (NA35 collaboration), \NPA{566}{1994}{35c}.

\bibitem{Gazdzicki95}
M.~Ga\'zdzicki et al. (NA35 collaboration), \NPA{590}{1995}{197c}.

\bibitem{Guenther95}
J.~G\"unther et al. (NA35 collaboration), \NPA{590}{1995}{487c}.

\bibitem{Murray95}
M.~Murray et al. (NA44 collaboration), AIP Conference Proceedings 
{\bf 340}, {\it Strangeness '95}, J.~Rafelski (eds.), AIP Press, 
1995, p.~162.

\bibitem{Santo94}
R.~Santo et al. (WA80 collaboration), \NPA{566}{1994}{61c}.

\bibitem{Lebedev94}
A.~Lebedev et al. (WA80 collaboration), \NPA{566}{1994}{355c}.

\bibitem{Albrecht95a}
R.~Albrecht et al. (WA80 collaboration), \PLB{361}{1995}{14}.

\bibitem{Albrecht95b}
R.~Albrecht et al. (WA80 collaboration), \PRL{76}{1996}{3506}

\bibitem{Agakichiev95}
  G.~Agakichiev et al.~ (CERES collaboration), \PRL{75}{1995}{1272}.

\bibitem{Venugopalan94}
  R.~Venugopalan, M.~Prakash, M.~Kataja, and P.V.~Ruuskanen,
  \NPA{566}{1994}{473c}.

\bibitem{Ornik89}
U.~Ornik, F.W.~Pottag, and R.M.~Weiner, \PRL{63}{1989}{2641}.

\bibitem{Schnedermann92}
E.~Schnedermann and U.~Heinz, \PRL{69}{1992}{2908};\\
E.~Schnedermann, J.~Sollfrank, and U.~Heinz, \PRC{48}{1994}{2468}.

\bibitem{Bolz92}
J.~Bolz, U.~Ornik, and R.M.~Weiner, \PRC{46}{1992}{2047}.

\bibitem{Katscher93}
U.~Katscher et al.~, \ZPA{346}{1993}{209}.

\bibitem{Gersdorff86}
H.~von Gersdorff, L.~McLerran, M.~Kataja, and P.V.~Ruuskanen,
\PRD{34}{1986}{794}.

\bibitem{Boris73}
J.P.~Boris and D.L.~Book, \JCP{11}{1973}{38};\\
D.L.~Book, J.P.~Boris and K.~Hain, \JCP{18}{1975}{248}.

\bibitem{Kataja88}
M.~Kataja, \ZPC{38}{1988}{419}.

\bibitem{Zalesak79}
S.T.~Zalesak, \JCP{31}{1979}{335}.

\bibitem{Rischke95a}
  D.H.~Rischke, S.~Bernard, and J.A.~Mahrun, \NPA{595}{1995}{346}.

\bibitem{karschQM95}
  F. Karsch, {\it Nucl. Phys.} {\bf A590} (1995) 367c.

\bibitem{Blum95}
 T.~Blum, S.~Gottlieb, L.~K\"arkk\"ainen, and D.~Toussaint,
 {\it Nucl.~Phys.} {\bf B} (Proc.~Suppl.) {\bf 42} (1995) 460.

\bibitem{BlaiOll86}J.-P. Blaizot and J. Y. Ollitrault, 
{\it Nucl.Phys.} {\bf A458} (1986) 745; D.H. Rischke and M. Gyulassy, 
{\it Nucl.Phys.} {\bf A597} (1996) 701.

\bibitem{RajThesis}R. Venugopalan, Ph. D. Thesis, SUNY Stony Brook, 
August 1992, Unpublished.

\bibitem{WVP}
  G. Welke, R. Venugopalan and M. Prakash, 
  {\it Phys. Lett.} {\bf 245} (1990) 137.

\bibitem{Venugopalan92}
  R.~Venugopalan and M.~Prakash, \NPA{546}{1992}{718}.

\bibitem{Hagedorn94}
  R.~Hagedorn and J.~Ranft, {\it Nuovo Cim.~Suppl.} {\bf 6} (1968) 
  169;\\
  R.~Hagedorn, in {\it Hot Hadronic Matter: Theory and Experiment},
  J.~Letessier et al. (Eds.), Plenum, New York, 1995, and
  references therein.

\bibitem{Letessier95}
  J.~Letessier, A.~Tounsi, U.~Heinz, J.~Sollfrank, and J.~Rafelski,
  \PRD{51}{1995}{3408}.

\bibitem{Kapusta89}
J.~Kapusta, ``Finite Temperature Field Theory'', 
Cambridge University Press, Cambridge, 1989.

\bibitem{Kapusta83a}
  J.I.~Kapusta and K.A.~Olive, \NPA{408}{1983}{478}.

\bibitem{Walecka74}
  J.D.~Walecka, \AP{83}{1974}{491}; B.D~Serot and J.D.~Walecka, 
  Advances in Nuclear Physics, Vol 16, ed. J.W. Negele and E. Vogt 
  (Plenum, NY, 1986);\\
  B.D.~Serot, Rep. Prog. Phys. {\bf 55} (1992) 1855.

\bibitem{Greiner}
 W.~Greiner, ``Thermodynamics and Statistical Mechanics'', 
 Springer Verlag, New York, 1995.


\bibitem{Rischke95b}
  D.H.~Rischke, Y.~P\"urs\"un, and J.A.~Mahrun, \NPA{595}{1995}{383}.

\bibitem{Bjorken83}
J.D.~Bjorken, \PRD{27}{1983}{140}.

\bibitem{Blobel74}
  V.~Blobel et al., \NPB{69}{1974}{454}.

\bibitem{Otterlund83}
  I.~Otterlund, \NPA{418}{1983}{98c}.

\bibitem{Woods54}
  R.D.~Woods and D.S.~Saxon, \PRv{95}{1954}{577}.

\bibitem{Goity89}
  J.L.~Goity and H.~Leutwyler, \PLB{228}{1989}{517}.

\bibitem{RajPrak93}
M.~Prakash, M.~Prakash, R.~Venugopalan and G.~Welke, 
\PR{227}{1993}{323}; \PRL{70}{1993}{1228}.

\bibitem{Schnedermann94}
  E.~Schnedermann and U.~Heinz, \PRC{50}{1994}{1675}.

\bibitem{Sorge95}
  H.~Sorge, \PLB{373}{1996}{16}

\bibitem{Cooper74}
  F.~Cooper and G.~Frye, \PRD{10}{1974}{186}.

\bibitem{Sollfrank91}
 J.~Sollfrank, P.~Koch, and U.~Heinz, \ZPC{52}{1991}{593}.

\bibitem{GaKap}
 C.~Gale and J.~Kapusta, \NPB{357}{1991}{65}.

\bibitem{BraPi90}
 E.~Braaten and R.D.~Pisarski, \NPB{337}{1990}{569}.

\bibitem{Kapusta91}
  J.~Kapusta, P.~Lichard, and D.~Seibert, \PRC{45}{1991}{2774}.

\bibitem{Baier92}
  R.~Baier, H.~Nakkagawa, A.~Ni\'egawa, and K.~Redlich,
  \ZPC{53}{1992}{433}.

\bibitem{Traxler95}
 C.~Traxler, H.~Vija, and M.~Thoma, \PLB{346}{1995}{329}.

\bibitem{Nadeau92}
  H.~Nadeau, J.~Kapusta, and P.~Lichard, \PRC{45}{1992}{3034};
  \PRC{47}{1993}{2426}.

\bibitem{Xiong92}
  L.~Xiong, E.~Shuryak, and G.E.~Brown, \PRD{46}{1992}{3798}.

\bibitem{Lichard} P.~Lichard and M.~Prakash, to be published.

\bibitem{Cleymans87}
   J.~Cleymans, J.~Fingberg, and K.~Redlich, \PRD{35}{1987}{2153}.

\bibitem{BrPiYa90}
 E.~Braaten, R.D.~Pisarski and T.-C.~Yuan, \PRL{64}{1990}{2242}.

\bibitem{AlRu92}
 T.~Altherr and P.V.~Ruuskanen, \NPB{380}{1992}{377}.

\bibitem{Gale94}
  C.~Gale and P.~Lichard, \PRD{49}{1994}{3338}.

\bibitem{Lichard94} P. Lichard, \PRD{49}{1994}{5812}.

\bibitem{Lichard96} P. Lichard, private communication.

\bibitem{Steele96}{J. Steele, H. Yamagishi, and I. Zahed }, 
\PLB { } {1996} {to be published}.

\bibitem{Srivastava94a}
D.K.~Srivastava, J.~Pan, V.~Emel'yanov, and C.~Gale, 
\PLB{329}{1994}{157}.

\bibitem{Srivastava96}
D.K.~Srivastava, B.~Sinha, and C.~Gale,
\PRC{53}{1996}{R567}.

\bibitem{Cassing95}
W.~Cassing, W.~Ehehalt, and C.M.~Ko, \PLB{363}{1995}{35}.

\bibitem{PDG94}
Particle Data Group, \PRD{50}{1994}{1173}.

\bibitem{Landsberg85}
 L.G.~Landsberg, \PR{128}{1985}{301}.

\bibitem{Huovinen96}
P.~Huovinen, M.Sc.~thesis, University of Jyv\"askyl\"a, 1996, 
unpublished.

\bibitem{Aguilar-Benitez91}
M.~Aguilar--Benitez et al. (NA27 collaboration), \ZPC{50}{1991}{405}.

\bibitem{Dumitru95}
A.~Dumitru et al., \ZPC{51}{1995}{2166}.

\bibitem{Srivastava94b}
D.K.~Srivastava and B.~Sinha, \PRL{73}{1994}{2421};\\
D.K.~Srivastava and B.~Sinha, \NPA{590}{1995}{507c}.

\bibitem{Shuryak94}
E.V.~Shuryak and L.~Xiong, \PLB{333}{1994}{316}.

\bibitem{Arbex95}
N.~Arbex, U.~Ornik, M.~Pl\"umer, A.~Timmermann, and R.M.~Weiner,
\PLB{354}{1995}{307}.

\bibitem{Tserruya} I.~Tserruya, private communication.

\bibitem{Koch93}
P.~Koch, \ZPC{57}{1993}{283}.

\bibitem{Li95}
G.Q.~Li, C.M.~Ko, and G.E.~Brown, \PRL{75}{1995}{4007};
G.Q.~Li, C.M.~Ko, and G.E.~Brown, \NPA{}{1996}{,~to be published.}

\bibitem{Kapusta95}
J.~Kapusta, D.~Kharzeev, and L.~McLerran, \PRD{53}{1996}{5028}

\bibitem{Redlich95}
K.~Redlich, J.~Cleymans, and V.V.~Goloviznin, University of Bielefeld
preprint, BI--TP--94--48.

\end{thebibliography}
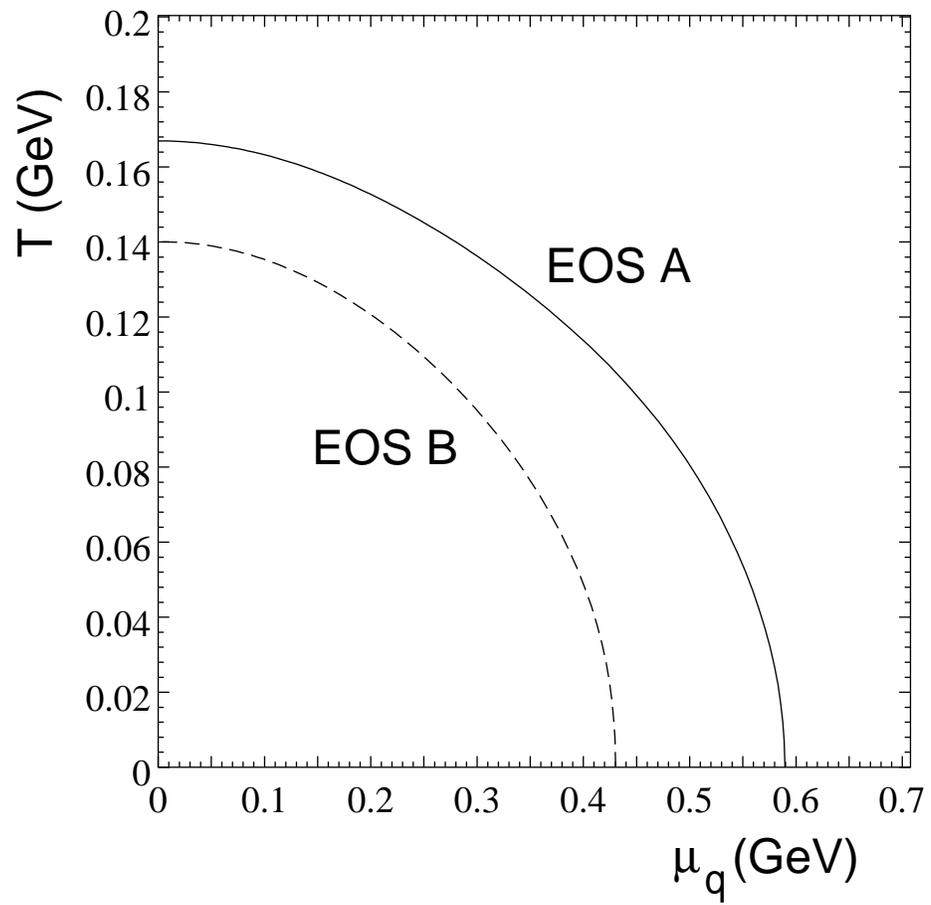
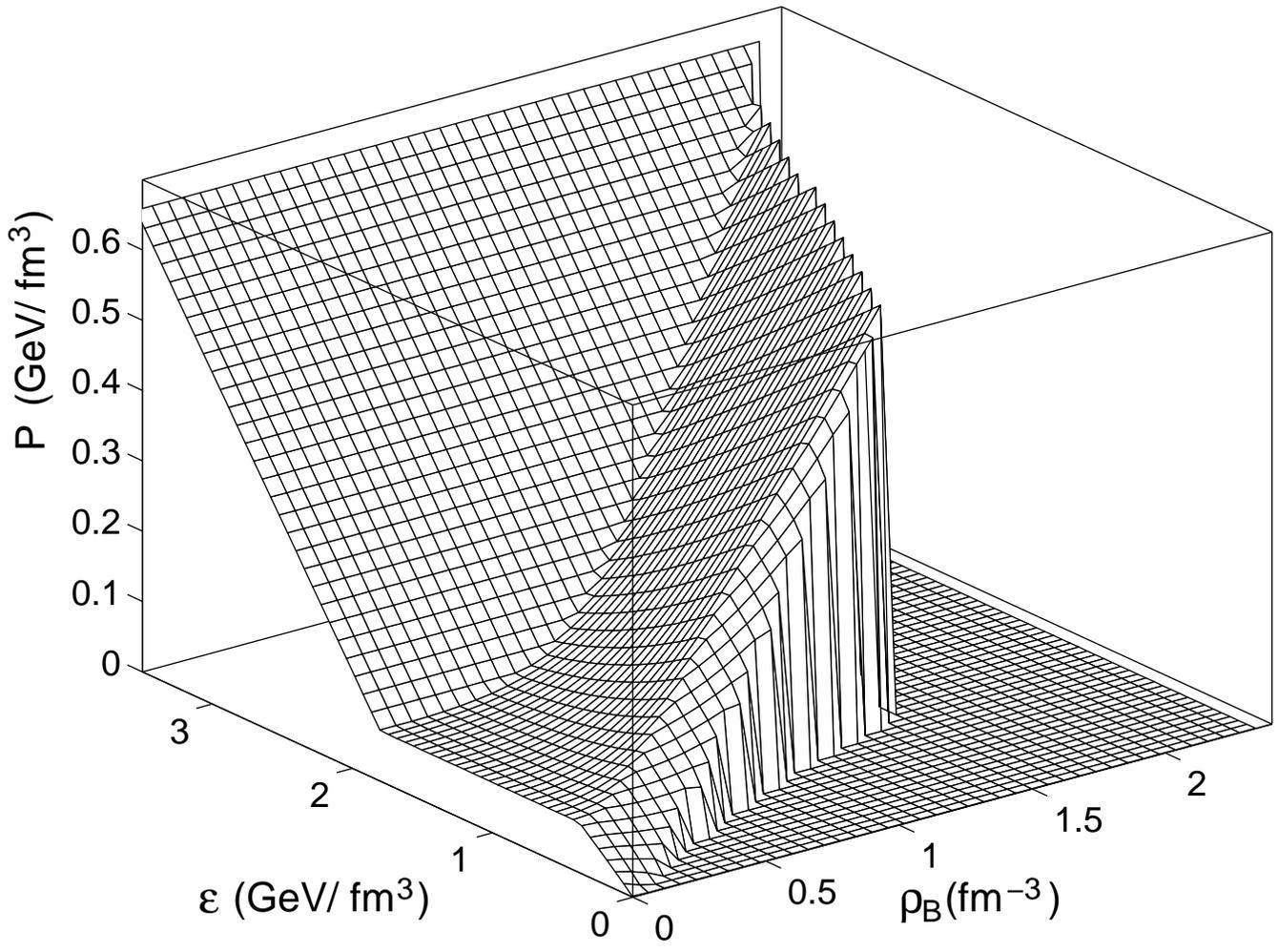
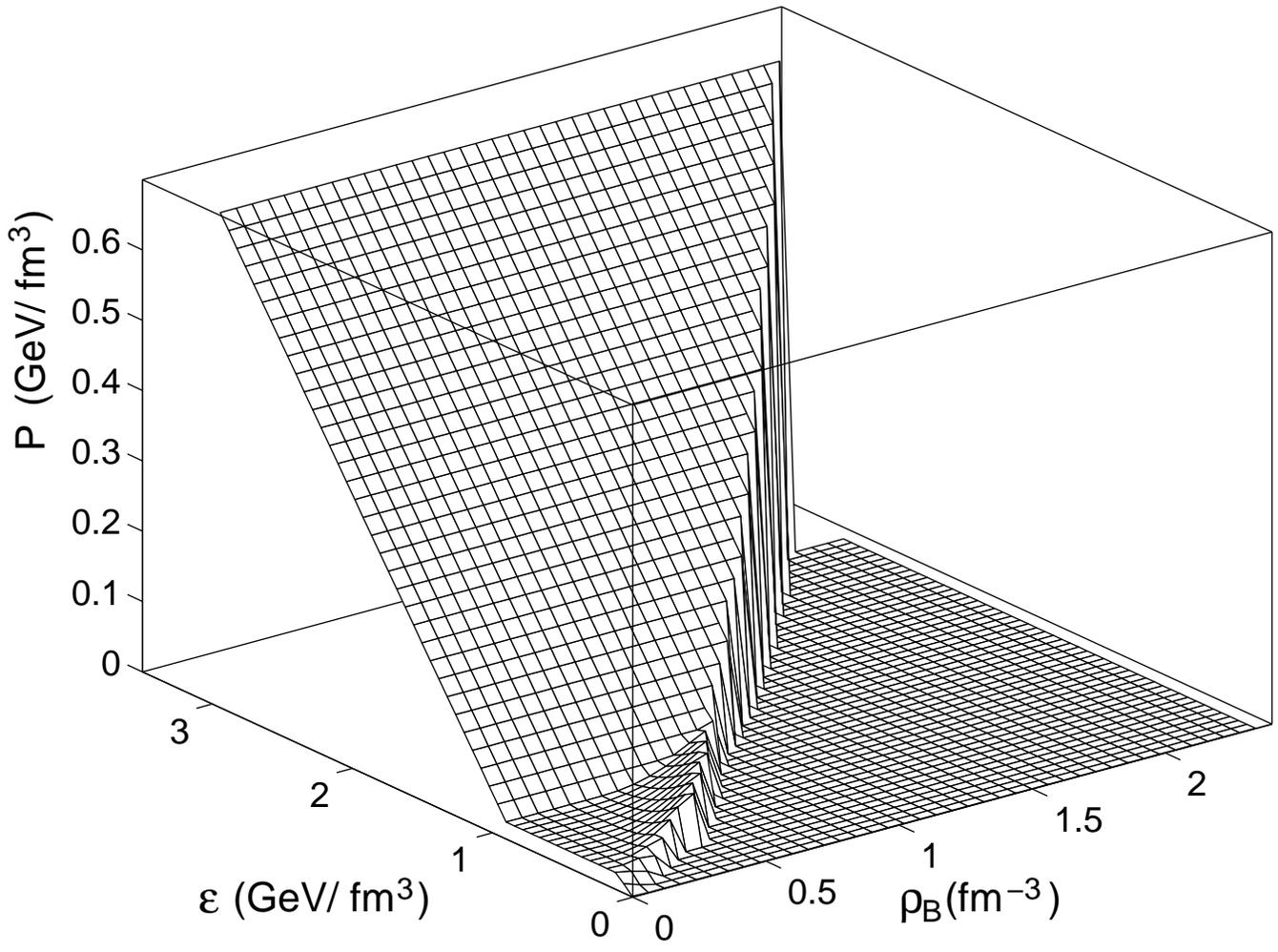
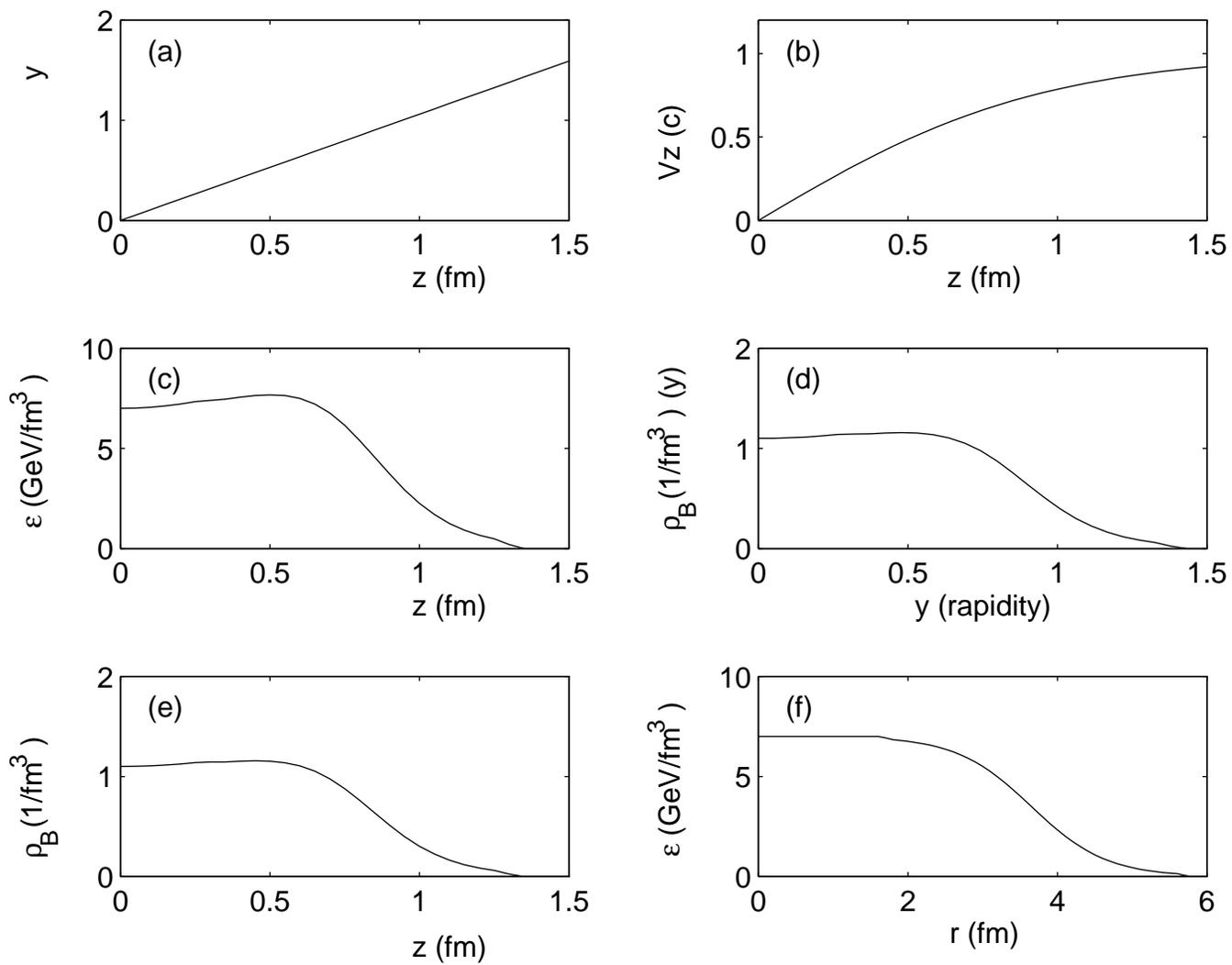
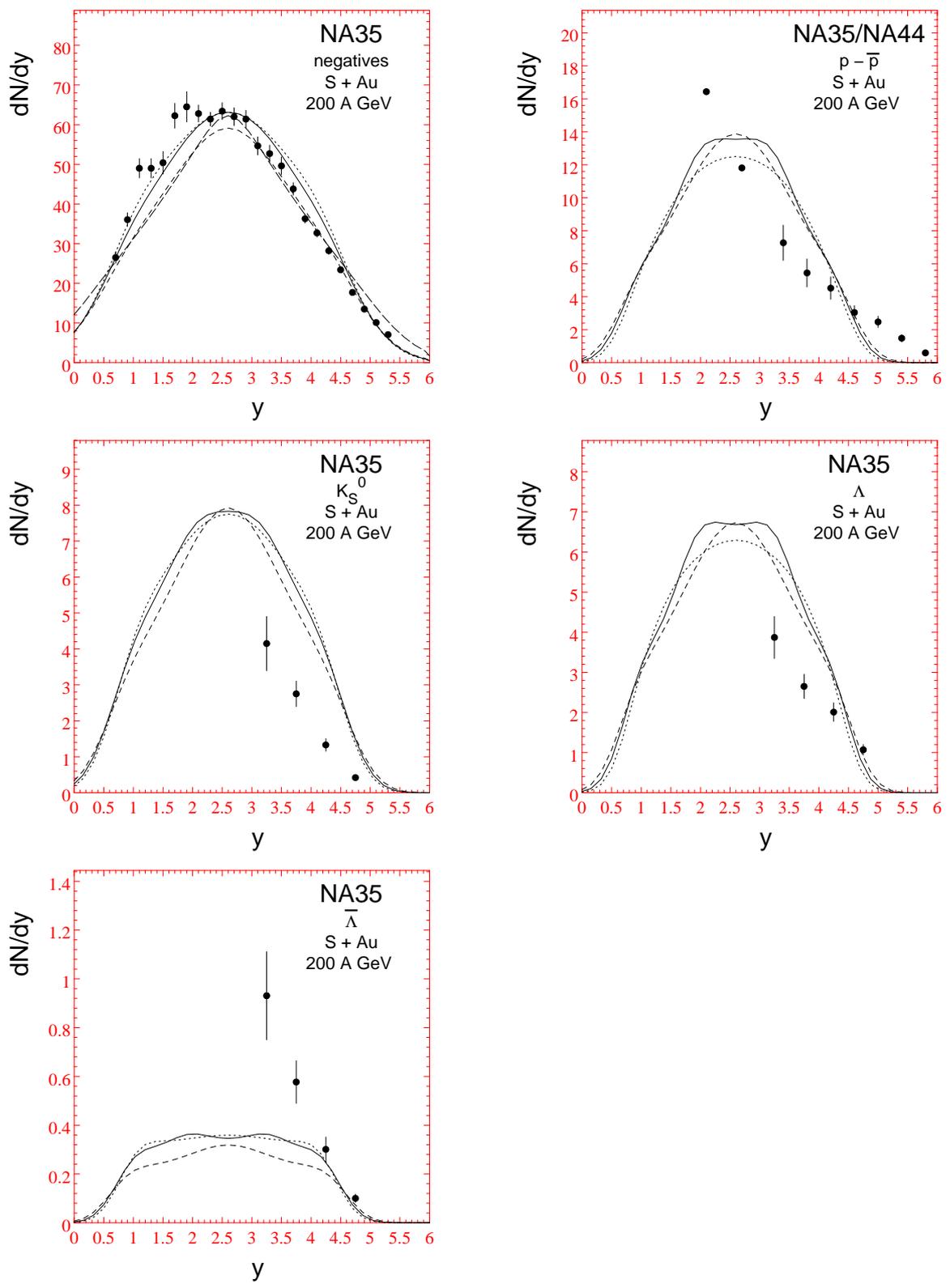
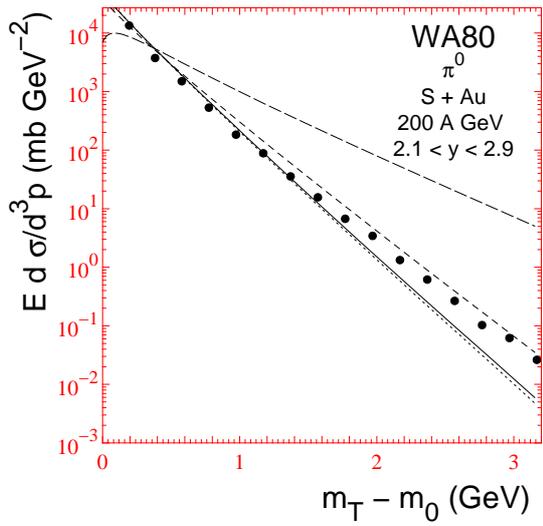
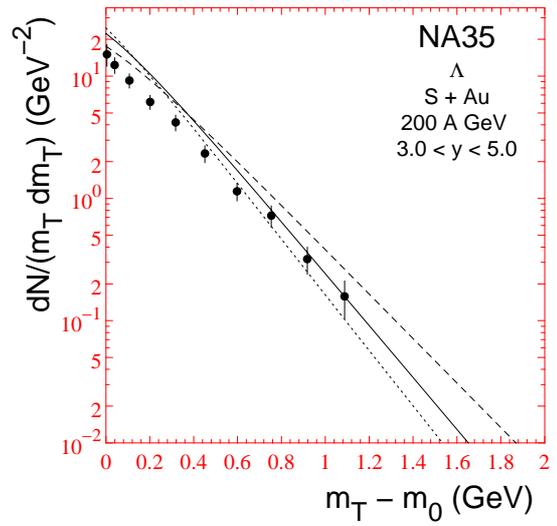
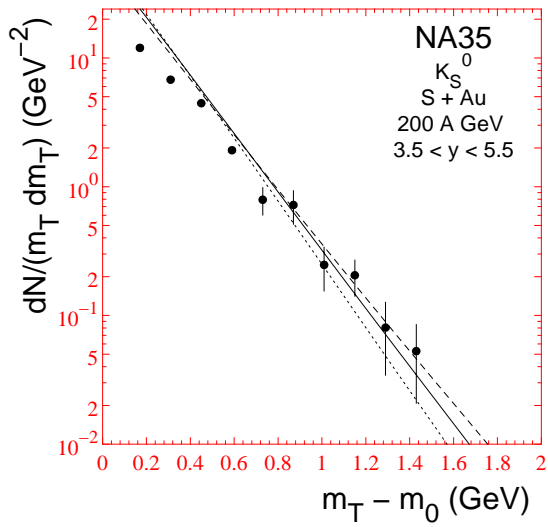
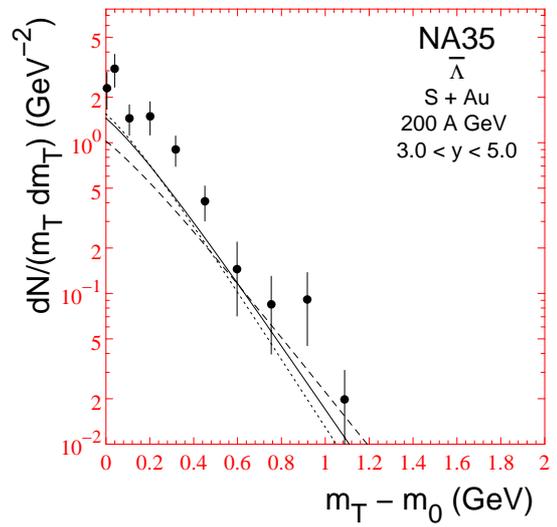
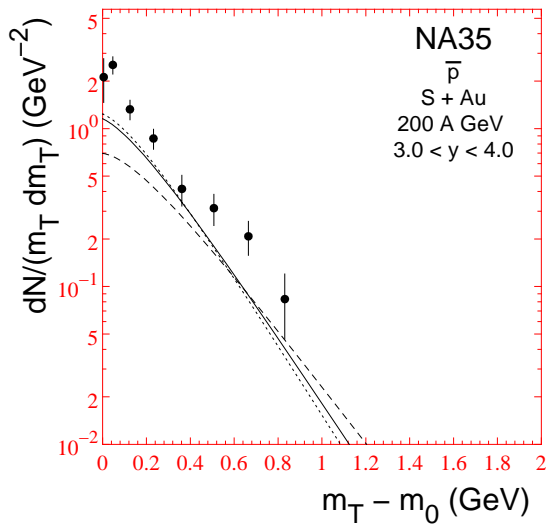
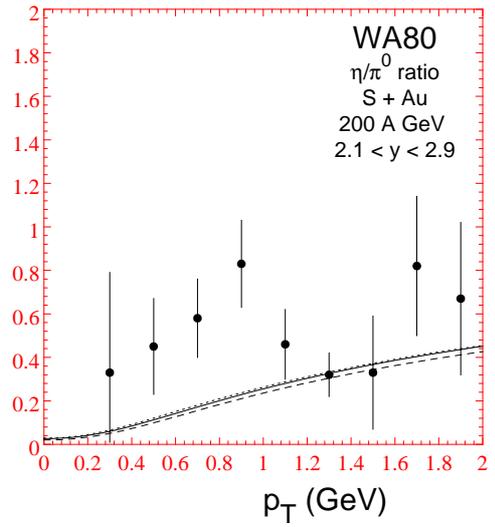
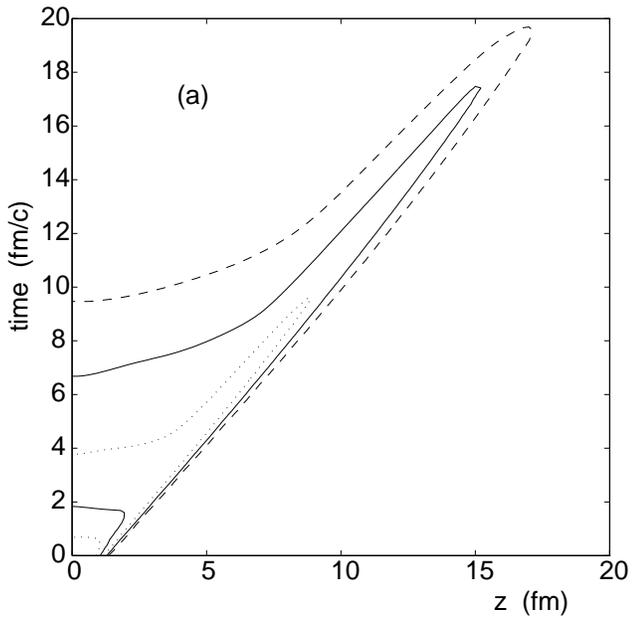
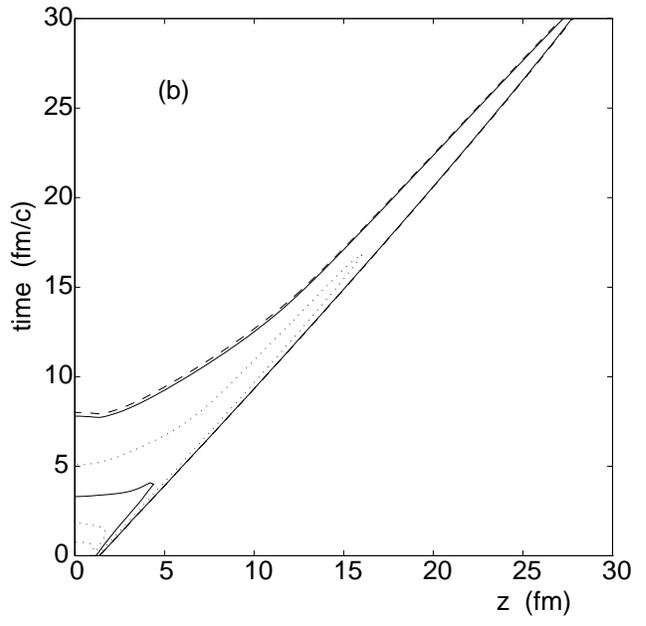
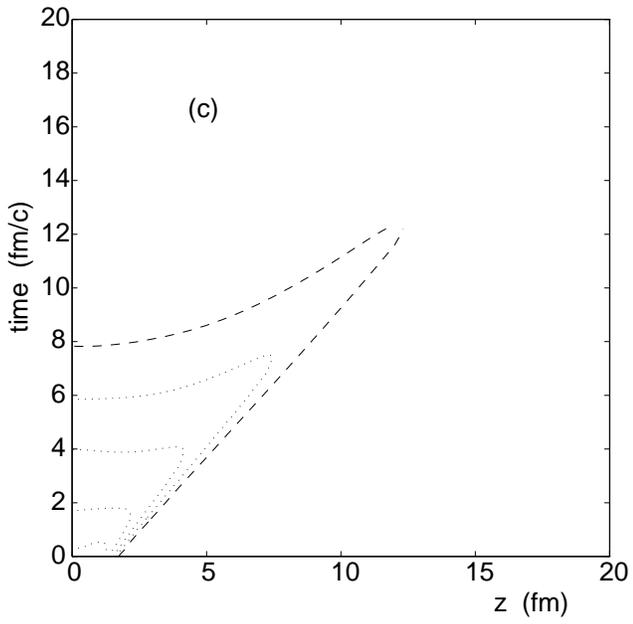
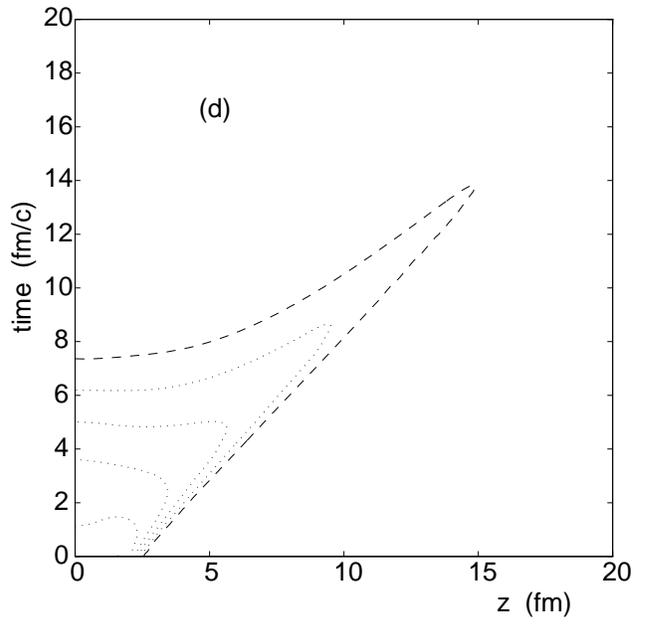
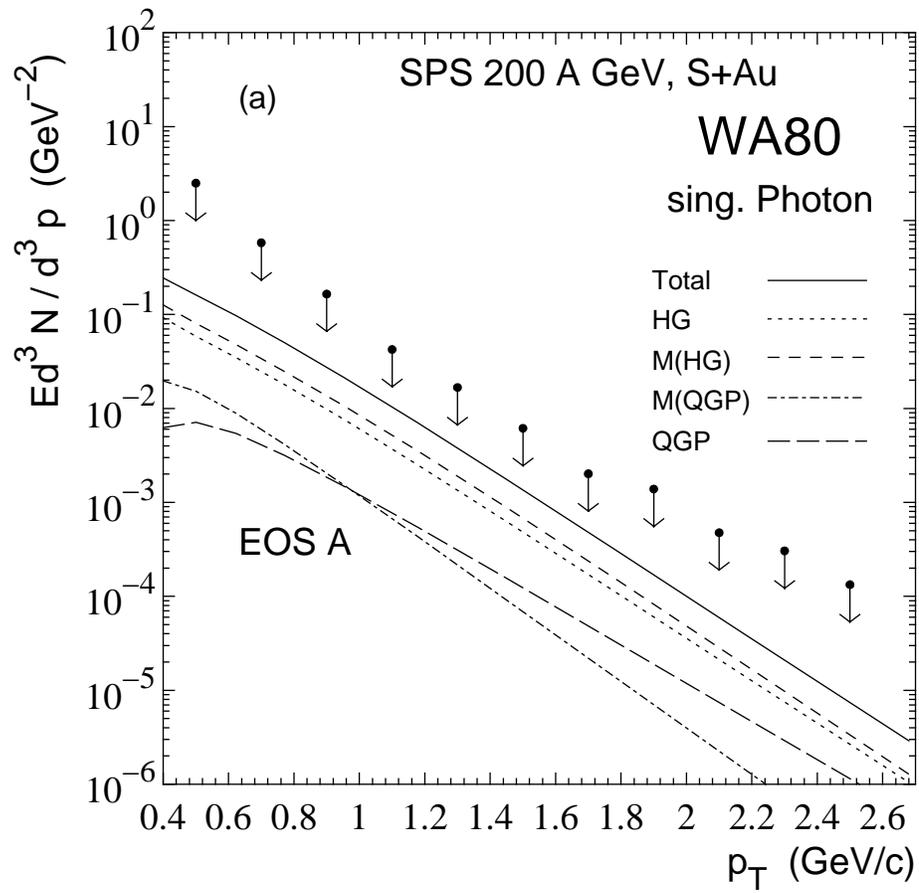
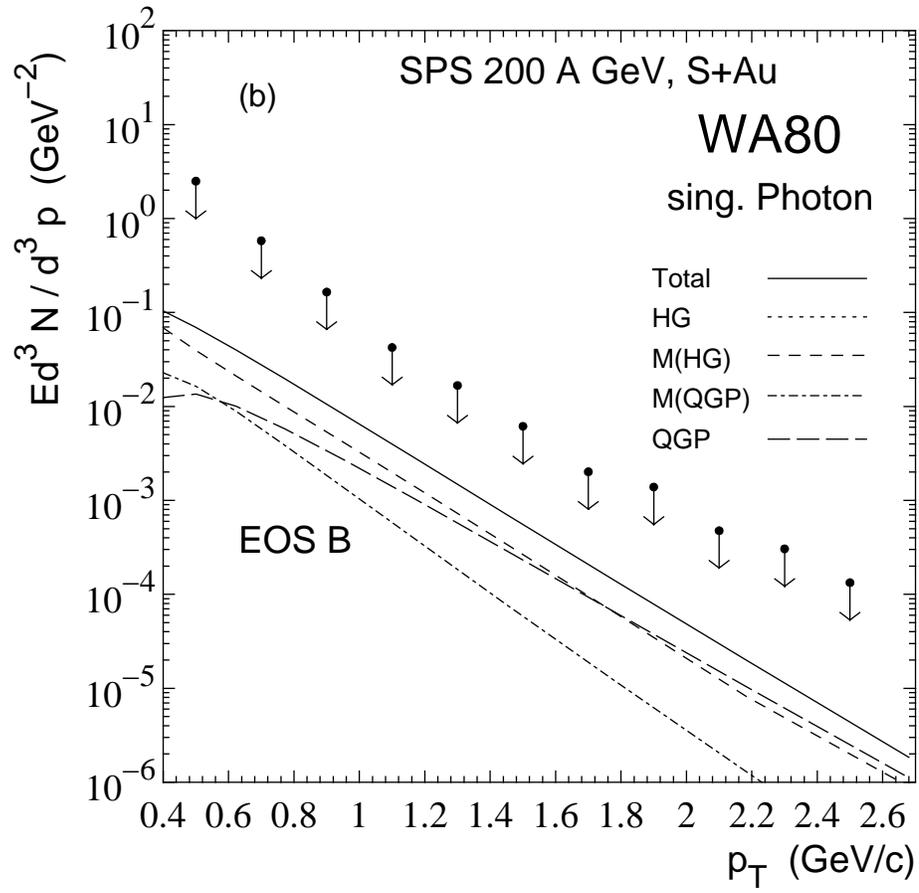
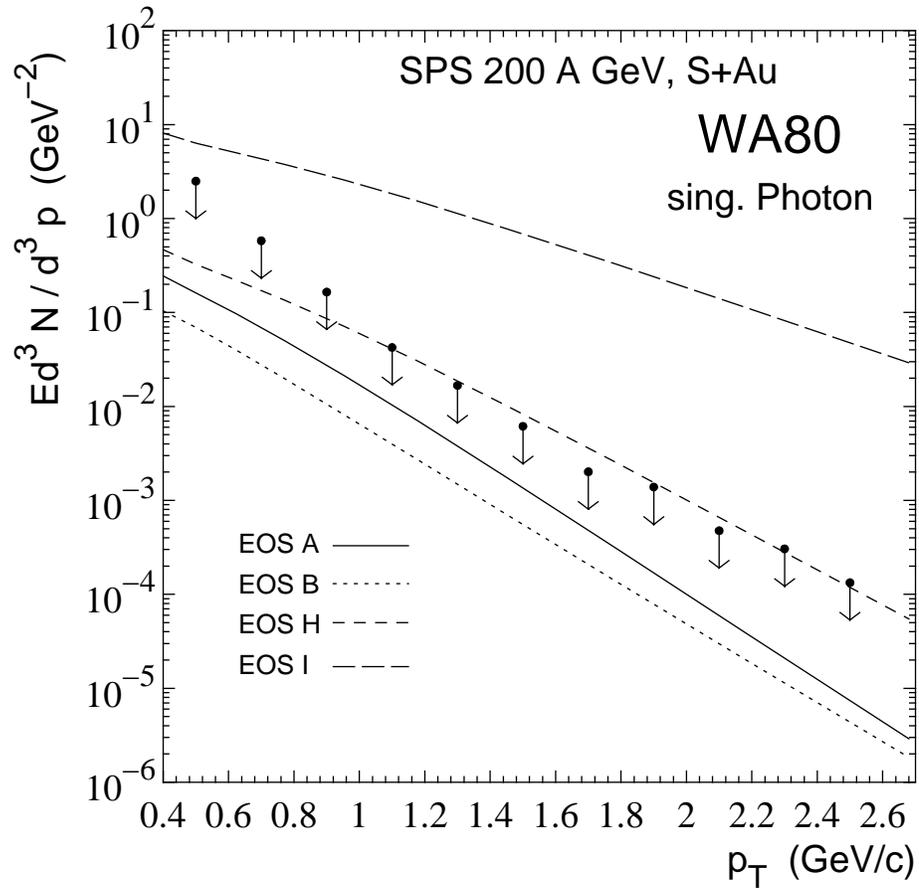
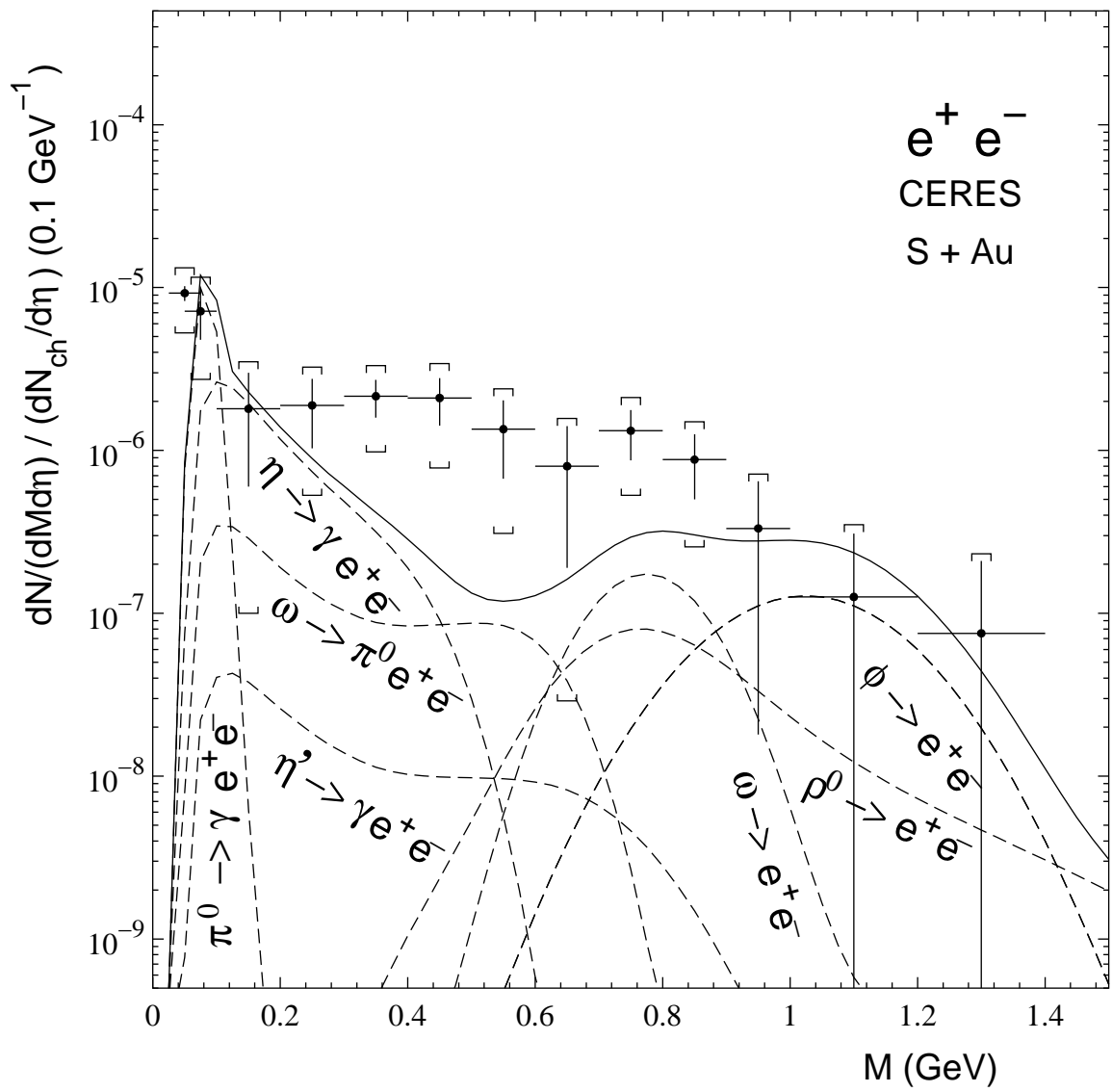
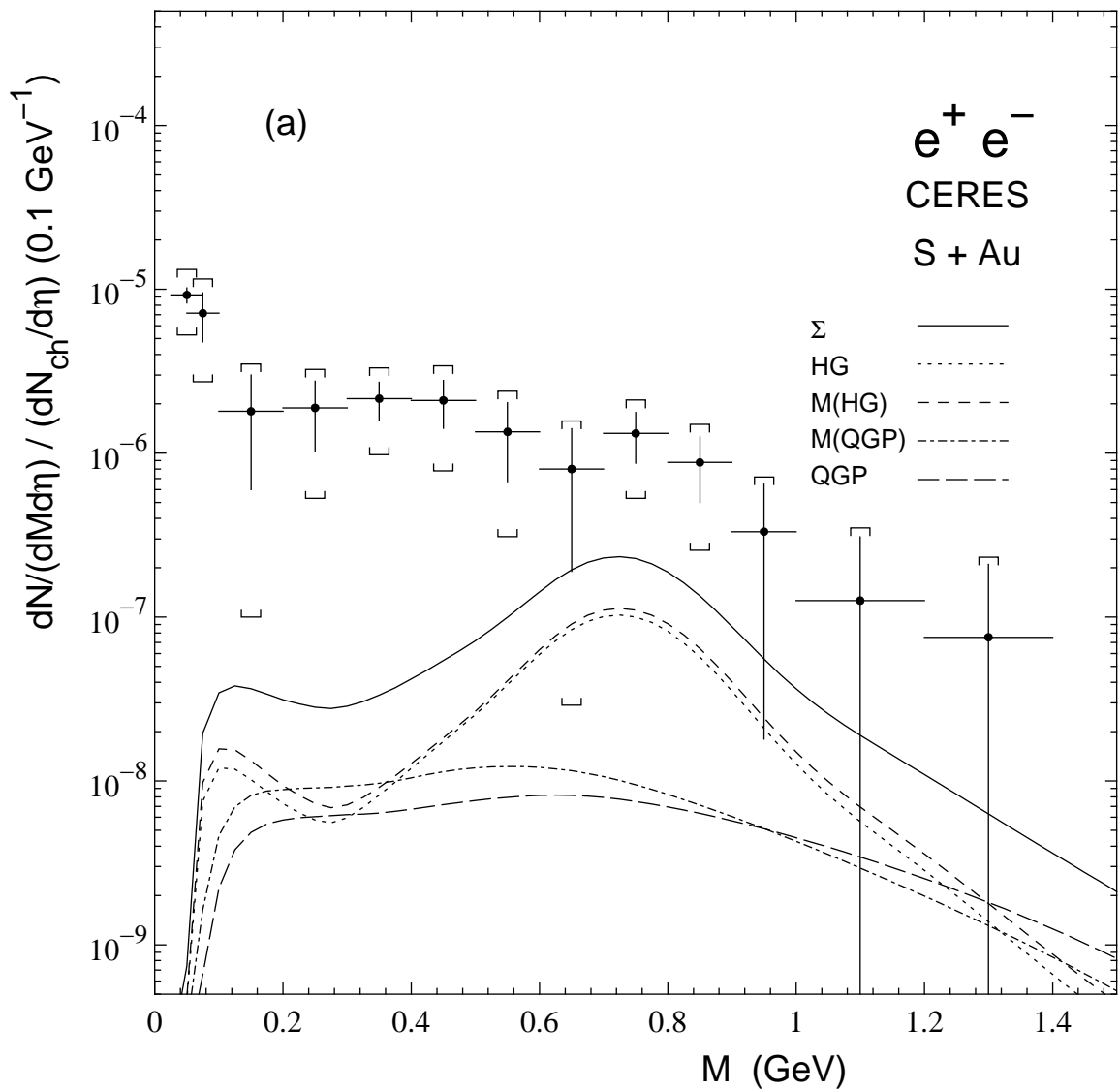
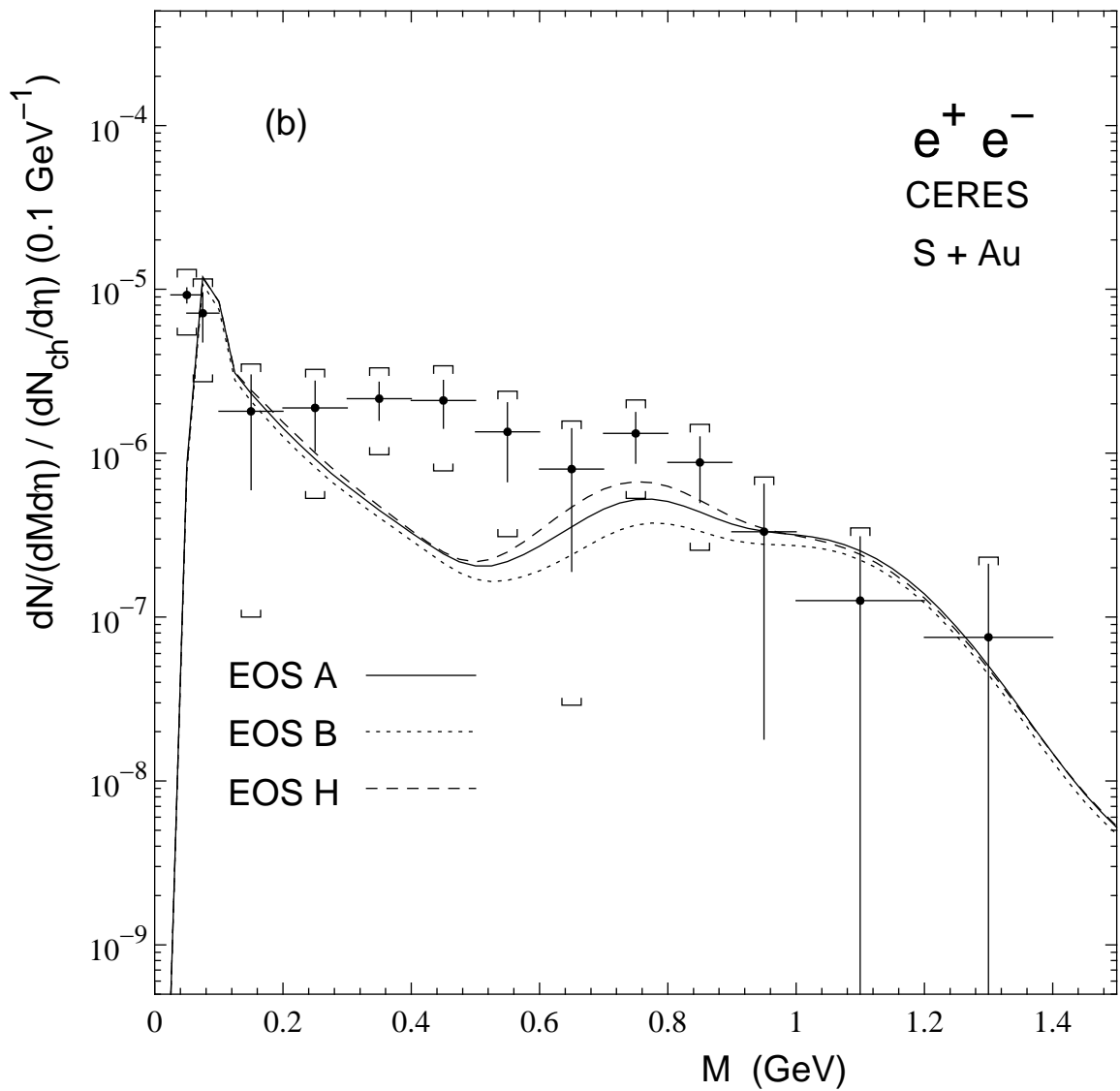
\end{document}